\documentclass{aa}  
\usepackage{graphicx}
\usepackage{xcolor}
\usepackage{amsmath}
\usepackage[separate-uncertainty=true]{siunitx}
\usepackage{enumitem}
\usepackage[options]{adjustbox}
\usepackage{txfonts}
\usepackage{hyperref}
\usepackage[nameinlink]{cleveref}

\hypersetup{
    colorlinks=true,
    citecolor=blue,
    linkcolor=purple,
    urlcolor=magenta,
}

\DeclareSIUnit{\parsec}{\mathrm{pc}}
\DeclareSIUnit{\gauss}{\mathrm{G}}
\DeclareSIUnit{\erg}{\mathrm{erg}}
\DeclareSIUnit{\year}{\mathrm{yr}}
\DeclareSIUnit[]\solarmass{\mathrm{M}_\odot}
\DeclareSIUnit[]\dyne{\mathrm{dyne}}

\creflabelformat{equation}{#2\textup{#1}#3}

\newcommand*\diff{\mathop{}\!\mathrm{d}}

\begin{document} 

   \title{Impact of Cosmic Ray Acceleration on the Early Evolution of Bow Shocks around Massive Runaway Stars}

   \author{
        K. Watanabe\inst{1}
        \and
        S. Walch\inst{2}
        \and
        T.-E. Rathjen\inst{2}
        \and
        J. Mackey\inst{3}
        \and
        P. C. N\"{u}rnberger\inst{2}
        \and
        P. Girichidis\inst{4}
    }

    \institute{
        Institut f\"{u}r Astroteilchenphysik,
        Karlsruhe Institute of Technology, 
        Hermann-von-Helmholtz-Platz 1, 
        76344 Eggenstein-Leopoldshafen, 
        Germany
        \and
        Universit\"{a}t zu K\"{o}ln, 
        I. Physikalisches Institut,
        Z\"{u}lpicher Str. 77,
        50937 K\"{o}ln,
        Germany
        \and
        Astronomy \& Astrophysics Section, School of Cosmic Physics, Dublin Institute for Advanced Studies, DIAS Dunsink Observatory, Dublin D15 XR2R, Ireland
        \and
        Universit\"{a}t Heidelberg, 
        Zentrum f\"{u}r Astronomie, 
        Institut f\"{u}r Theoretische Astrophysik, 
        Albert-Ueberle-Str. 2, 
        D-69120 Heidelberg, 
        Germany
    }


    
    \abstract
    {
    Bow shocks generated from the interaction of winds from massive runaway stars with the interstellar medium have been shown to be prominent particle accelerators through recent $\gamma$-ray and radio synchrotron observations. They are ideal candidates to explore the spatial- and time-dependence of particle acceleration due to their axisymmetric morphology and short dynamical timescales ($\sim$ 100 kyr).
    }
    {
    Current studies on the particle acceleration from bow shocks investigate the cosmic ray (CR) transport only in the test-particle limit. We aim to bridge this gap by conducting ideal cosmic ray magnetohydrodynamic (CRMHD) simulations in the advection-diffusion limit that accelerate CRs at spatially resolved shocks at each timestep. We qualitatively compare with current observations through the expected $\gamma$-ray and synchrotron emission.
    }
    {
    We perform 3D ideal CRMHD simulations with the Eulerian grid-based code FLASH, where stellar winds are injected through tabulated wind velocities and mass loss rates. We implement a gradient-based shock detection algorithm to resolve the shocked regions where the CRs are injected dynamically. Simulations are performed for different values of the CR diffusion coefficient and star velocities within an ISM-like environment up to 500 kyr to showcase the impact of dynamical CR injection on the early evolution of the wind-driven bow shock. With a simplified spectral model in post-processing, we calculate the expected upper limits of $\gamma$-ray and synchrotron emission and compare with those from current observations.
    }
    {
    We observe that variations of CR diffusion rates can strongly dictate the morphology of the bow shock and the overall $\gamma$-ray and radio synchrotron luminosity due to the balance between the CR injection efficiency and diffusion. Our results are underestimated compared to those from current observations, even with an overestimated particle acceleration, primarily due to the weaker wind kinematic luminosities used and a lack of a stellar magnetic field in our model. 
   }
   {
   The morphology of wind-driven bow shocks and their non-thermal emission can be substantially affected by CR acceleration and diffusion, if there is efficient particle acceleration at shocks. 
   }


    \maketitle


    \section{Introduction}
    \label{sec:introduction}
    Massive stars have been shown to be prominent particle accelerators and thus factories for Galactic cosmic rays \citep{Aharonian2018}. The fast-moving winds generated from stellar radiation apply ram pressure onto the ambient interstellar medium (ISM), forming a stellar wind bubble \citep{Weaver1977}. The shocks may have high enough Mach numbers to induce particle acceleration \citep{Blandford1978}, which can be readily observed through $\gamma$-ray or radio synchrotron emission \citep{DelRom12}. Non-thermal emission from wind-driven shocks has been observed from wind nebulae surrounding Wolf-Rayet stars (e.g. \citealp{Prajapati2019}) and young stellar clusters \citep{TavSabPia09, Fenech2015, Yang2018, Muno2006, HESS2022_Westerlund1}. The wind bubbles generated from such clusters have also been shown to contribute up to 10\% of the total Galactic cosmic ray population \citep{Peron2024}. Magnetohydrodynamic (MHD) simulations further indicate that the cluster-wind termination shock can accelerate particles which are consistent with current $\gamma$-ray observations \citep{Haerer2025}.\par 

Massive stars moving with respect to the ISM are similarly predicted to be sites for particle acceleration. Following the detection of radio synchrotron emission from the bow shock of the massive runaway star BD+43\,3654 \citep{BenRomMar10}, theoretical models were developed to predict potential high-energy radiation through pion decay and/or inverse-Compton radiation \citep{DelRom12}.
Searches for this high-energy non-thermal radiation at keV \citep{ToaOskIgn17}, GeV \citep{SchAckBue14}, and TeV \citep{HESS2018_Bowshocks} energies did not achieve any detections, with upper limits excluding the more optimistic theoretical models.
More recent multi-zone calculations \citep{DelBosMul18} and post-processing methods applied to 2D hydrodynamic simulations \citep{delValle2018} show that while X-ray detection of synchrotron or inverse Compton radiation is very challenging, GeV and TeV emission may not be too far below current observational upper limits.
Recently 3D MHD simulations have emerged as a valuable tool to investigate thermal \citep{Mackey2025}, non-thermal \citep{delValle2025}, and both \citep{Martinez:2023} emission from bow shocks, but so far cosmic rays were only included in the test-particle limit in a post-processing step.
One-zone models \citep{MouMacCar22} and 1D MHD simulations of stellar wind bubbles \citep{Meyer2024} have also shown that the re-acceleration of pre-existing cosmic rays is consistent with the non-thermal emission observed from BD+43\,3654.
The relatively short timescale ($\sim \SI{100}{\kilo\year}$) of bow shocks means that they should be close to a time-stationary state, and so they are a potentially valuable laboratory for studying spatially resolved particle acceleration and transport processes (advection and diffusion) in the ISM.
In particular, the bow shock of $\zeta$ Ophiuchi \citep{GreMacKav22}, at a distance of only 135\,pc and with an angular scale of $\approx5$\,arcmin, is an excellent candidate \citep{DelRom12, DelBosMul18}.\par 

Radio observations are advancing rapidly with the advent of broad-band and wide-field observations at GHz frequencies.
\citet{MouMacCar22} detected a non-thermal spectrum from the bow shocks of BD+43\,3654, confirming earlier work \citep{BenRomMar10, BenDelHal21}, and also from the Bubble Nebula, the bow shock driven by the runaway star BD+60\,2522.
\citet{VanSaiMoh22} detected radio emission from a number of bow shocks with \textsc{MeerKAT}, albeit without spectral information to distinguish between thermal and non-thermal radiation. \citet{vandenEijnden2024} have also detected synchrotron radio emission without a gamma-ray counterpart from LS2335 through the Australian Square Kilometre Array Pathfinder (ASKAP) survey.
With this rapid development in sensitivity at radio frequencies and the ongoing construction of the Cherenkov Telescope Array Observatory \citep{CTAO2013}, it is important to make fully self-consistent 3D calculations of MHD together with particle acceleration, transport, and radiation, to identify the most promising candidates for high-energy targeted observations. \par

In this work, we simulate the 3D evolution of wind-driven bow shocks generated from massive runaway stars through cosmic-ray magnetohydrodynamic (CRMHD) simulations. We implement a framework that injects cosmic rays (CRs) through spatially resolved shocks at each timestep, assuming that the CRs are accelerated through diffusive shock acceleration (DSA) \citep[e.g.][]{Krymskii1977, Axford1978, Bell1978a, Bell1978b, Blandford1978, Caprioli2014}. Through these simulations, we investigate the impact of dynamic shock-driven CR acceleration on the early evolution of the bow shock. We also analyse the observable emission through the synchrotron and gamma-ray processes induced by CRs, and match with current observations. \par 

In \Cref{sec:numerical_impl}, we describe the numerical framework used to simulate massive runaway stars, in particular highlighting the implementation of the shock detection and cosmic ray injection algorithm required for CR acceleration in shocks.
\Cref{sec:runawaystars} shows the application of our numerical framework to simulate massive runaway stars with varying environmental conditions. In particular, we highlight the impact of dynamic CR injection to the short time and spatial scales of the bow shock. 
In \Cref{sec:observations}, we compare our simulated results with current $\gamma$-ray and radio synchrotron observations, where we use simplified models to convert our hydrodynamic parameters to observable quantities. We discuss possible shortcomings and improvements of our current simulation framework in \Cref{sec:discussion} and conclude in \Cref{sec:conclusions}.

    \section{Numerical Implementation}
    \label{sec:numerical_impl}
    In recent years, the development of astrophysical simulation frameworks that include the CR transport equation to the standard 3D MHD equations have progressed significantly, in particular due to the importance of CRs in driving galactic winds on galactic scales \citep{Naab2017, Zweibel2017} and stellar outflows in the ISM \citep{Girichidis2016, Girichidis2018}. Nevertheless, due to the 6D (spatial and momentum) phase space of the CR spectrum, it is non-trivial to combine both equations while limiting the computational cost.  As such, current astrophysical frameworks focus on including CRs in the advection-diffusion limit through a `grey' approach \citep[e.g.][]{Dubois2019, Pfrommer2017}. Here, the CRs are treated as an additional fluid, and their properties are integrated over the full CR momentum space. In this way, the global impact of CRs (through advection or diffusion) can be captured. Works that consider a spectrally resolved CRMHD equation such as \texttt{CREST} \citep{Winner2019} or \texttt{CRSPEC} \citep{Yang2017, Girichidis2020}, which allow momentum-dependent energy loss and diffusion to be considered, have also started to emerge \citep[see][for a recent review]{Ruszkowski2023}. \par 

\subsection{The CRMHD Equations} \label{subsec:numerical_methods}

In our work, we use the Eulerian grid-based MHD code \texttt{FLASH} in version 4 \citep{Fryxell2000, Dubey2008} that utilises adaptive mesh refinement (AMR) to save computational effort in regions of low interest. We solve the combined CRMHD equations using an adapted version of \texttt{FLASH} that includes physical processes such as supernova and stellar feedback, chemical networks, cosmic rays, and radiative transfer \citep{ Walch2015, Gatto2017, Girichidis2018, Rathjen2021, Rathjen2023, Rathjen2025, Brugaletta2025}. The ideal MHD equations are solved using a directionally split HLL3R finite-volume scheme \citep{Bouchut2010, Waagan2011}, where the divergence of the magnetic field is treated using a parabolic diffusion method following \cite{Marder1987}. The CR transport equations are solved following the numerical implementation as described in \cite{Girichidis2018}, where the CRs are treated as a separate monoenergetic fluid integrated over all CR energies. \par 

The combined system of equations that we solve numerically is as such:
\begin{align}
    &\frac{\partial \rho}{\partial t} + \vec{\nabla} \cdot (\rho \vec{u}) = \dot{\rho}_\mathrm{wind}, \\
    &\frac{\partial \left(\rho \vec{u}\right)}{\partial t} + \vec{\nabla} \cdot \left(\rho \vec{u}\vec{u}^T - \frac{\vec{B} \, \vec{B}^T}{4\pi}\right) + \vec{\nabla} P = \dot{\vec{q}}_\mathrm{wind}, \\
    &\frac{\partial \epsilon}{\partial t} + \vec{\nabla} \cdot \left[(\epsilon + P) \vec{u} - \frac{\vec{B}(\vec{B} \cdot \vec{u})}{4\pi}\right] \\
    & \qquad \qquad \qquad = \vec{\nabla} \cdot \left[\mathcal{K} \vec{\nabla}  \epsilon_\mathrm{CR}\right] + \Lambda_\mathrm{cool} + \Lambda_\mathrm{hadr} + Q_\mathrm{CR} + Q_\mathrm{wind}, \\
    &\frac{\partial \vec{B}}{\partial t} - \vec{\nabla} \times (\vec{u} \times \vec{B}) = 0, \\
    &\frac{\partial \epsilon_\mathrm{CR}}{\partial t} + \vec{\nabla} \cdot (\epsilon_\mathrm{CR} \vec{u}) = - P_\mathrm{CR} (\vec{\nabla} \cdot \vec{u}) + \vec{\nabla} \cdot \left[\mathcal{K} \vec{\nabla}  \epsilon_\mathrm{CR}\right] + Q_\mathrm{CR} + \Lambda_\mathrm{hadr} ,
    \label{eq:cr-mhd}
\end{align}
where $\rho$ is the gas density, $\vec{u}$ is the fluid velocity, and $\vec{B}$ is the magnetic field. The total energy density $\epsilon$ contains the kinetic, thermal, magnetic, and CR contributions
\begin{equation}
    \epsilon = \rho |\vec{u}|^2 / 2 + \epsilon_\mathrm{th} + B^2 / 8\pi + \epsilon_\mathrm{CR} 
    \label{eq:en_tot}
\end{equation}
and the total pressure $P$ similarly contains thermal, magnetic, and CR contributions:
\begin{equation}
    P = P_\mathrm{th} + B^2 / 8\pi  + P_\mathrm{CR}.
    \label{eq:ptot}
\end{equation}
The thermal and CR pressure can be expressed in terms of energy densities through their closure relation, given as:

\begin{equation}
    P_\mathrm{th} = (\gamma_\mathrm{th} - 1) \epsilon_\mathrm{th} \quad\mathrm{and}\quad P_\mathrm{CR} = (\gamma_\mathrm{CR} - 1) \epsilon_\mathrm{CR},
    \label{eq:eos}
\end{equation}
where $\gamma_\mathrm{th}$ and $\gamma_\mathrm{CR}$ are the thermal and CR adiabatic indices, respectively. \par 

To describe stellar wind feedback, we follow the prescription used in \cite{Gatto2017} where we inject the wind feedback in the form of kinetic momentum, $\dot{\vec{q}}_\mathrm{wind}$, and subsequently through the kinetic energy, $Q_\mathrm{wind}$, and mass, $\dot{\rho}_\mathrm{wind}$. The injected momentum at each timestep is calculated by determining the wind luminosity $L_\mathrm{wind}$ through tabulated values of the mass loss rate $\dot{M}$ and terminal velocity $v_\mathrm{wind}$ for each stellar mass. These tables are obtained from stellar evolution tracks from the zero-age main sequence star to the Wolf-Rayet / pre-supernova phase \citep{Ekstroem2012}. The winds are injected uniformly within a spherically symmetric region with a radius of $r_{w,\mathrm{inj}} = 4 \times \Delta x$ where $\Delta x$ is the smallest cell size corresponding to the maximum resolution. \par 

To reduce the computational complexity of the problem, we do not determine cooling rates from tracing the chemical evolution of species; rather we describe the radiative cooling $\Lambda_\mathrm{cool} = \Lambda_\mathrm{cool}(T, Z)$ using the semi-analytical function of \cite{Koyama2002} for $T \leq \SI{1e4}{\kelvin}$ and cooling tables from \cite{Plewa1995} for $T > \SI{1e4}{\kelvin}$. We assume optically thin cooling throughout the simulation. \par

The CRs in our work are described in a simplified advection-diffusion approximation where we assume that the CRs are scattered sufficiently enough from small-scale background turbulence. The CR diffusion is described anisotropically with constant diffusion coefficients parallel and perpendicular to the magnetic field ($\kappa_\parallel$ and $\kappa_\perp$, respectively), expressed as such:
\begin{equation}
    \mathcal{K} := \kappa_{ij} = \kappa_\perp \delta_{ij} + (\kappa_\parallel - \kappa_\perp) b_i b_j, 
    \label{eq:crdifftensor}
\end{equation}
where $b_i = B_i / |\vec{B}|$ denotes the component of the magnetic field in each direction. We assume that the diffusion is strongly anisotropic and predominantly runs along the field lines such that $\kappa_\perp \sim 0.01 \kappa_\parallel$ \citep{Nava2013}. \par 

In this work we also assume that the CR particles consist only of protons and are ultra-relativistic with a constant CR adiabatic index of $\gamma_\mathrm{CR} = 4/3$. The hadronic losses, $\Lambda_\mathrm{hadr}$, in our framework are calculated using the nucleon density and the CR energy density in each cell, following the prescription of \cite{Girichidis2018}. Coulomb losses are not considered since modeling such a loss mechanism is only viable with a spectrally resolved model. \par 

While \cite{Girichidis2018} assumes that all CR injection occurs within a certain injection radius, we implement a new on-the-fly framework within \texttt{FLASH} that self-consistently determines the CR injection energy $Q_\mathrm{CR}$ from resolving the shocked region at each timestep. In the upcoming sections, we elaborate on the procedure and present tests to show the validity of our implementation.

\subsection{CR shock acceleration}  \label{subsec:crshock_acc}

The acceleration mechanism of CRs in astrophysical scenarios, in particular at the highest energies, have not yet been fully understood. Despite this, the regularity of the CR spectrum below $\sim$ PeV energies indicate that there is a standard paradigm that describes CR acceleration within the Galaxy. First-order Fermi acceleration, or Diffusive Shock Acceleration (DSA), has been put forth as such a mechanism, where particles are repeatedly scattered across the shock, gaining energy proportional to the shock velocity. 
This mechanism has been studied both theoretically \citep[e.g.][]{Krymskii1977, Axford1978, Bell1978a, Bell1978b, Blandford1978, Drury1983}, and also through the use of hybrid particle-in-cell (PIC) simulations, where they calculate the collisionless interactions between individual electrons and ions near the shock within a magnetised medium \citep[see][for a recent review]{Pohl2020}. Through this paradigm, the CR injection follows a power-law in momentum, $f_\mathrm{inj}(p) \propto p^{-\alpha_\mathrm{inj}}$, where $\alpha_\mathrm{inj} = 4$. Non-linearities from DSA, caused by varying shock compressions experienced by different particles near the shock, have been investigated, showing variations of $\alpha_\mathrm{inj} > 4$ \citep[e.g.][]{Ellison1996,  Jones1991, Malkov2001, Amato2005, Amato2006, Kang2005, Kang2006,Caprioli2008, Caprioli2009b, Caprioli2020, Diesing2021}. The currents generated from the ions also lead to magnetic field amplification \citep{Bell2004}, which is crucial for the self-regulation of efficient DSA at weaker shocks \citep[e.g.][]{Bamba2005, Voelk2005, Parizot2006, Morlino2012, Ressler2014}. \par 

The treatment for non-linear DSA within a 3D CRMHD framework is challenging, requiring the detection of subshocks very close to the shock that will locally modify the injection index, further modifying the CR momentum distribution per cell. As such, we implement CR acceleration at resolved shocks following the standard DSA mechanism. In order to achieve this, we need to first identify the shocked regions at each timestep, which are acceleration sites for CRs, and calculate the energy injected for CR acceleration. 

\subsubsection{Shock Detection}  \label{subsec:shock_det}

Determining and resolving the shocks from MHD simulations at each timestep is a challenge. Due to numerical diffusion, the shock is spread over a few cells, which makes it non-trivial to identify the pre-shock and post-shock states. The finite size of each cell also adds an additional complexity when identifying the shock normal (unit vector directed perpendicularly to the shock). Nevertheless, as mentioned in the previous section, it is necessary to resolve the shocked regions to not only locate where the CRs are injected but also to evaluate the Mach number $\mathcal{M}$ and magnetic obliquity $\theta_B$, which are used to calculate the injected CR energy within each cell. \par 

Our on-the-fly shock detection framework follows that of \cite{Pfrommer2017}, which is adapted from \cite{Schaal2015} for a composite fluid of thermal gas and CRs. We briefly summarise the algorithm here, highlighting parts adapted for our implementation. Here, we denote pre- and post-shock quantities with subscripts $1$ and $2$, respectively. \par 

The pre-shock Mach number, $\mathcal{M}_1$, for a composite thermal + CR gas can be obtained through the Rankine-Hugoniot jump conditions which also consider the CR pressure and energy density into account :
\begin{align}
    \mathcal{M}_1^2 = \frac{1}{\gamma_{\mathrm{eff}, 1}} \frac{y_s \mathcal{C}}{\mathcal{C} - \left[(\gamma_1 + 1) + (\gamma_1 - 1) y_s\right](\gamma_2 - 1)}; \\
    \mathcal{C} = \left[(\gamma_2 + 1)y_s + (\gamma_2 - 1)\right](\gamma_1 - 1)
    \label{eq:mach_cr}
\end{align}
where $y_s = P_{\mathrm{tot},2} / P_{\mathrm{tot},1}$ is the total pressure jump across the shock and $\gamma_i = P_i /  \epsilon_{\mathrm{int},i} + 1$ is the pre-/post-adiabatic index. The pre-shock effective adiabatic index, which characterises the mixing between the thermal and CR gas, is described by
\begin{equation}
    \gamma_{\mathrm{eff}, 1} = \left(\frac{\diff \log P}{\diff \log \rho} \right)_S = \frac{\gamma_\mathrm{th} P_{\mathrm{th}, 1} + \gamma_\mathrm{CR} P_{\mathrm{CR}, 1}}{P_{\mathrm{tot}, 1}}
    \label{eq:effgamma}
\end{equation}
which is evaluated at constant entropy $S$ \citep{Pfrommer2006}. \par

To identify the shocked regions, we first determine the shock normal in each cell by using the normalised pseudo-temperature gradient:
\begin{equation}
    \hat{n}_s = -\frac{\vec{\nabla} \tilde{T}}{|\vec{\nabla} \tilde{T}|} .
    \label{eq:shock_dir}
\end{equation}
As the CR pressure can contribute to the total pressure depending on the particular simulation setup, we use the pseudo-temperature, $\tilde{T}$, which instead takes into account both thermal and CR pressures, i.e. 
\begin{equation}
    \tilde{T} = \frac{\mu m_p (P_\mathrm{th} + P_\mathrm{CR})}{\rho}.
    \label{eq:pseudotemp}
\end{equation}
Here, $\mu = 0.61$ is the mean molecular weight for ionized gas and $m_p$ is the proton mass. The gradient is then calculated using a second-order central differencing scheme in each direction. \par 

We then identify the `shock zone' within our simulation, which consists of initial shock candidates defined on a cell-to-cell basis. The region is determined if the following criteria are met:
\begin{enumerate}[label=(\emph{\roman*})]
    \item the flow is converging ($\vec{\nabla} \cdot \vec{u} < 0$)
    \item  $\vec{\nabla} \tilde{T} \cdot \vec{\nabla} \rho > 0$ (thus filtering out any contact discontinuities)
    \item $\mathcal{M}_1 > \mathcal{M}_\mathrm{min}$ (to remove any shock-like features generated from numerical noise).
\end{enumerate}
As we cannot identify the pre- and post-shock regions at this step, we use the quantities from neighbouring cells to calculate the pre-shock Mach number to identify the shock zone. Following \cite{Pfrommer2017}, we set the minimum Mach number $\mathcal{M}_\mathrm{min} = 1.3$. \par 

\begin{figure}[!h]
    \centering
    \resizebox{\hsize}{!}{\includegraphics{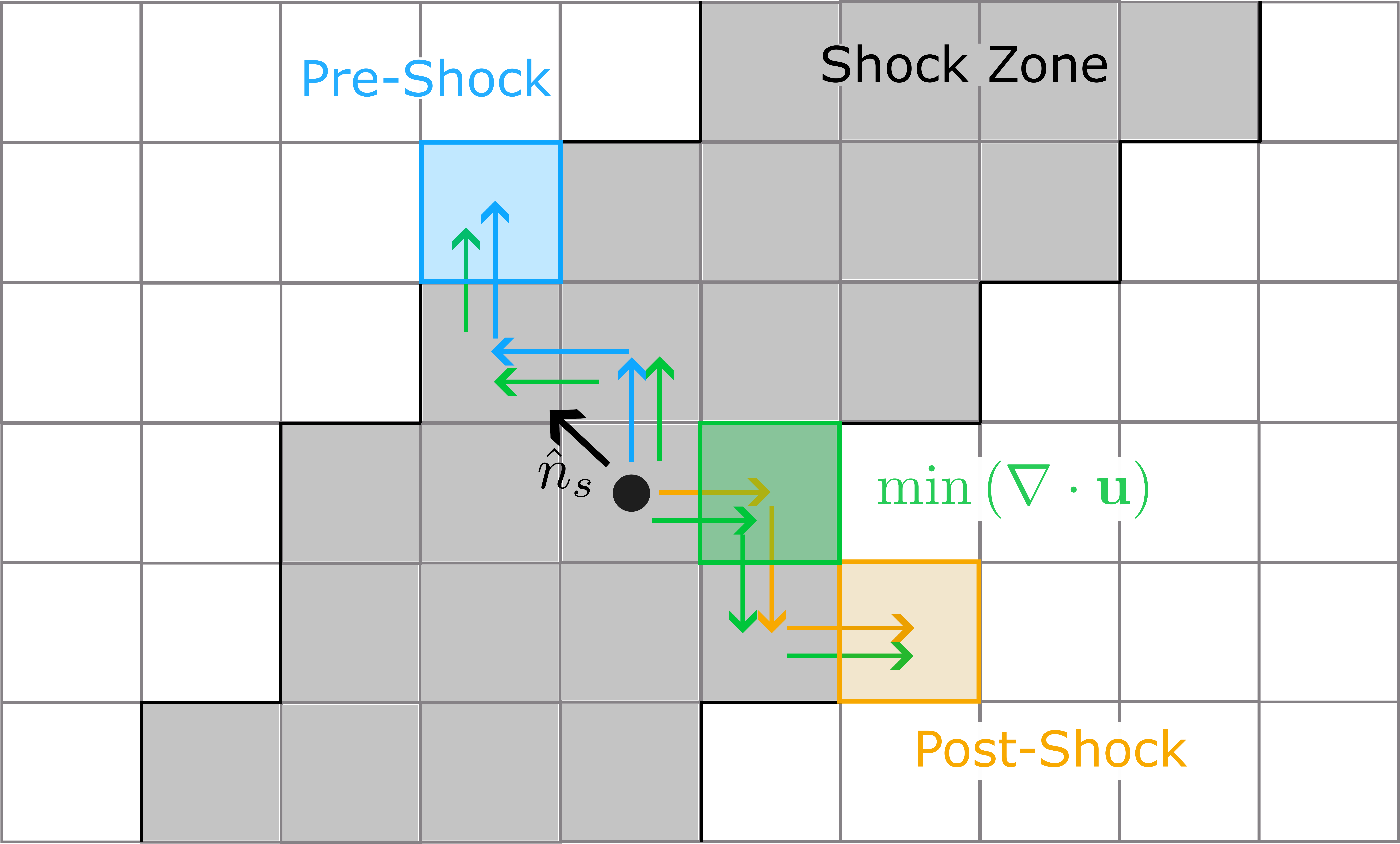}}
    \caption{Schematic of the implementation of the determination of the pre-shock (blue), post-shock (orange), and shock surface cells (green) for a shock candidate cell (black circle) within the shock zone. }
    \label{fig:prepostshock_schematic}
\end{figure}

We then determine for each shock candidate the corresponding cell of the pre-shock, the post-shock and the shock surface. Hereby, we first traverse the grid opposite the shock direction and record the location of the post-shock cell once it has exited the `shock zone'. Similarly, we determine the location of the pre-shock cell by moving along the shock direction. To determine the cell containing the shock surface we identify the cell with the lowest velocity divergence while we perform both pre-shock and post-shock searches. \Cref{fig:prepostshock_schematic} shows a schematic of the search algorithm. \par 

In this algorithm, we perform shocked cell searches within each block, described by a cube with eight cells in each dimension. While neighbouring blocks cannot be accessed directly, we use the guard cells, which store a local copy of the neighbouring blocks, four cells deep on each face of the cube. In this way, even if the pre-/post-shocked cells are located outside the active block, they can still be identified. The maximum number of cells used for this search is set to $N_\mathrm{sh, max} = 4$ to prevent our search algorithm from exceeding the number of cells in the local block (including the guard cells). \par 

We note that while we incorporate magnetic fields within our simulations, we only detect hydrodynamic shocks within this algorithm and leave the detection of MHD shocks to future works (see e.g. \citealt{Lehmann2016} for a post-processing shock detection module for MHD shocks).\par

\subsubsection{CR injection}  \label{subsec:crinj}

We now use the identified pre-shock, post-shock, and shock surface cells to calculate relevant shock parameters, which in particular are used to compute the CR injection energy. Similar to the previous section, we mainly follow the prescription from \cite{Pfrommer2017}, while adapting several components for our implementation in \texttt{FLASH}. \par 

We first calculate the pre-shock Mach number $\mathcal{M}_1$ using \Cref{eq:mach_cr} using the pre-shock and post-shock quantities, evaluated at the identified pre- and post-shock cells, respectively. We then calculate the rate at which the total energy is dissipated, $\dot{E}_\mathrm{diss}$, which is used to determine the energy used to accelerate CRs. Since the kinetic energy at the shock is dissipated into thermal and CR energies, we calculate the dissipated energy density by using the pre- and post-shock thermal and CR energy densities:
\begin{equation}
    \epsilon_\mathrm{diss} = (\epsilon_\mathrm{th, 2} - \epsilon_\mathrm{th,1} x_s^{\gamma_\mathrm{th}}) + (\epsilon_\mathrm{CR, 2} - \epsilon_\mathrm{CR,1} x_s^{\gamma_\mathrm{CR}}),
    \label{eq:diss_energy_density}
\end{equation} 
where the pre-shock energy densities are corrected for the adiabatic compression of the gas using the density jump $x_s = \rho_2 / \rho_1$. \par 
The dissipated energy flux, $f_\mathrm{diss}$, is then calculated using the post-shock velocity $v_2$ with the calculated dissipated energy density:
\begin{equation}
    f_\mathrm{diss} = \epsilon_\mathrm{diss} v_2 = \epsilon_\mathrm{diss} \frac{\mathcal{M}_1 c_1}{x_s},
    \label{eq:ediss_flux}
\end{equation}
where the post-shock velocity can be expressed in terms of the pre-shock Mach number $\mathcal{M}_1$, the pre-shock sound speed $c_1^2 = \gamma_\mathrm{eff} P_{\mathrm{tot},1} / \rho_1$ and the density jump $x_s$. \par 

\begin{figure}[!h]
    \centering
    \resizebox{\hsize}{!}{\includegraphics{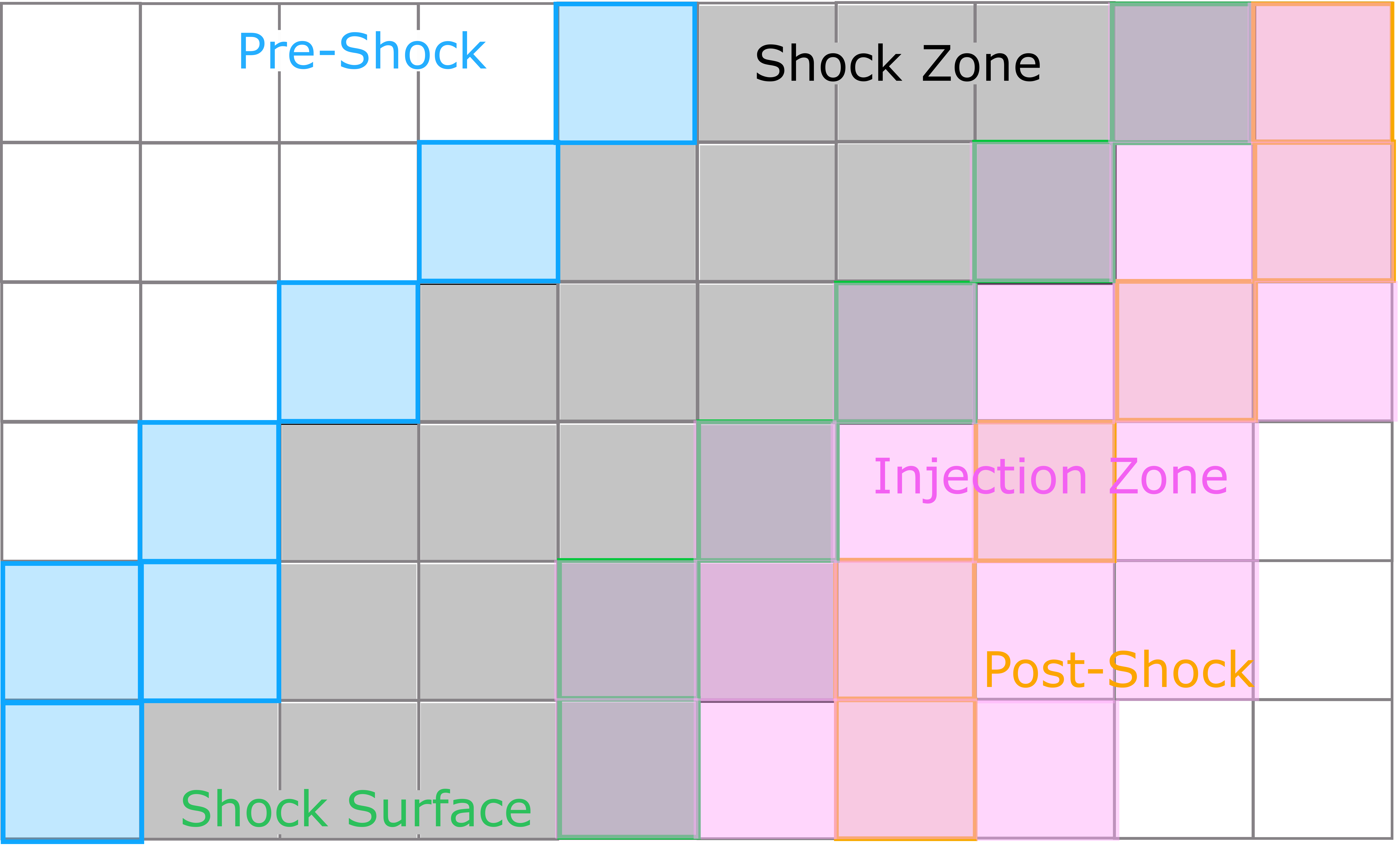}}
    \caption{Schematic of the implementation of CR injection within our framework. The `injection zone' (in pink), the numerically broadened shock surface used for energy injection, starts from the shock surface cells (green) and extends beyond the post-shock cells (yellow) up to a user-defined width $N_\mathrm{inj, max} = 4$. The pre-shock cells (blue) and the shock zone as defined in \Cref{fig:prepostshock_schematic} are also shown. }
    \label{fig:crinj_schematic}
\end{figure}

Instead of dissipating energy directly at the shock surface, we dissipate the energy in the `injection zone' which describes a numerically broadened shock surface used for energy injection. In our prescription, we define the injection zone from the determined shock surface cell (labelled as $s$) to a user-defined width $N_\mathrm{inj, max} = 4$. We note that the choice of this width is crucial to accurately model the dissipated energy, as decreasing this width can overestimate the amount of energy dissipated from the shock. This is highlighted in \Cref{fig:crinj_schematic}. \par 

To appropriately normalise for the numerically broadened shock, we follow the prescription from \cite{Pfrommer2017}:
\begin{equation}
    a_{\mathrm{inj}, l} = \frac{E_{\mathrm{int}, l} - E_{\mathrm{int, pre}}}{\sum\limits_{l^\prime = s}^{s + N_\mathrm{inj, max}} (E_{\mathrm{int}, l^\prime} - E_{\mathrm{int, pre}})},
    \label{eq:ediss_prefac}
\end{equation}
where $l \in \{s, s + N_\mathrm{inj, max}\}$ is used to label the cells within the injection zone and $E_{\mathrm{int}, l} = (\epsilon_{\mathrm{th}, l} + \epsilon_{\mathrm{CR}, l}) V_l$ is the total internal energy in each cell of volume $V_l$. The pre-shock internal energy, $E_{\mathrm{int, pre}}$, is computed similarly. \par 

The dissipated energy rate for each cell in the injection zone is then expressed as 
\begin{equation}
    \dot{E}_{\mathrm{diss}, l} = f_\mathrm{diss} A_\mathrm{shock} a_{\mathrm{inj}, l} ,
    \label{eq:ediss_dot}
\end{equation}
where $A_\mathrm{shock}$ is the projected surface area of a single cell along the shock direction. \par 

The energy injected to accelerate CRs is determined by using the dissipated energy rate, $\dot{E}_\mathrm{diss}$, in each cell within the injection zone with the CR acceleration efficiency $\eta = \eta(\mathcal{M}_1, \theta_B)$:
\begin{equation}
    E_{\mathrm{inj}, l} = \eta(\mathcal{M}_1, \theta_B) \dot{E}_{\mathrm{diss}, l} \Delta t 
    \label{eq:einj}
\end{equation}
where $\Delta t$ is the discretised timestep within our simulation and $\cos \theta_B = \vec{B} \cdot \hat{n}_s / |\vec{B}|$ is the cosine of the pre-shock magnetic obliquity. Separately from the Mach number criterion in the shock detection implementation, we apply a threshold Mach number $\mathcal{M}_{1, \mathrm{CR}} = 3.0$ to limit the injection of cosmic rays to strong shocks.  \par

A general description of the CR acceleration efficiency in astrophysical shocks is an unsolved problem, although progress has been made using hybrid particle-in-cell simulations \citep[e.g., ][]{Caprioli2014, Gupta2025}. Therefore, we adopt a semi-analytical approach through some commonly used and physically motivated prescriptions for the acceleration efficiency as a function of the shock Mach number and magnetic obliquity below. We first factorise the dependence of the Mach number and magnetic obliquity on the CR acceleration efficiency, i.e. $\eta(\mathcal{M}_1, \theta_B) = \xi(\mathcal{M}_1) \zeta(\theta_B)$. The dependence on the pre-shock Mach number $\mathcal{M}_1$ is modelled using a fitting function following \cite{Kang2007} that considers the treatment of both with and without the existence of presiding CRs within each cell (see \Cref{app:mach_eq} for the complete functional form). As this function overestimates the acceleration efficiency at large Mach numbers, we normalise the function such that the maximum value of this function is $\eta_\mathrm{max} = 0.2$ \citep{Caprioli2014, Kang2013}. \par

\begin{figure}[!h]
    \centering
    \resizebox{\hsize}{!}{\includegraphics{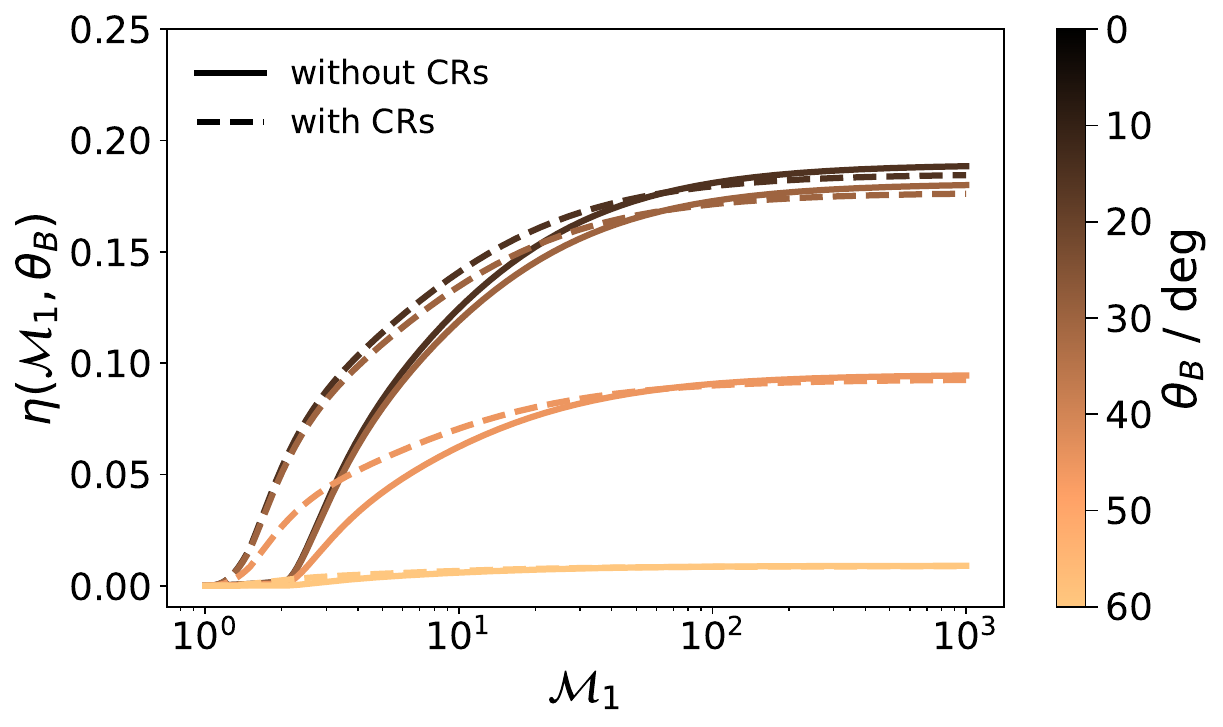}}
    \caption{Dependence of the CR acceleration efficiency $\eta(\mathcal{M}_1, \theta_B)$ on the pre-shock Mach number $\mathcal{M}_1$ and pre-shock magnetic obliquity $\theta_B$ at 15, 30, 45, and 60 degrees. Here, we use the prescription from \cite{Kang2007} and \cite{Pais2018} for the Mach number and magnetic obliquity dependence, respectively. The dashed and solid lines indicate the efficiency calculated with and without pre-existing CRs, respectively.}
    \label{fig:CRacceleff}
\end{figure}

To model the dependence with the pre-shock magnetic obliquity, we follow the semi-analytical prescription as described in \cite{Pais2018}:
\begin{equation}
    \zeta(\theta_B) = \frac{1}{2} \left[\tanh \left(\frac{\theta_\mathrm{crit} - \theta_B}{\delta_\theta}\right) + 1 \right],
    \label{eq:craccel_mgob}
\end{equation}
where $\theta_\mathrm{crit} = \pi / 4$ and $\delta_\theta = \pi / 18$. \Cref{fig:CRacceleff} shows the dependence of the CR acceleration efficiency on both $\mathcal{M}_1$ and $\theta_B$ using the prescription above. We do not take into account the dependence of the acceleration efficiency on the CR-to-thermal pressure ratio (see e.g. \citealt{Dubois2019} for an implementation of the dependency to the CR-to-thermal pressure ratio).

\subsection{Test of shock detection and numerical CR shock acceleration}  \label{subsec:crshock_test}

To verify our implementation, we perform a series of standardised (magneto-)hydrodynamic tests and compare with existing analytical solutions of such tests. Note that all tests are performed in computational units, thus we omit units when describing the setup of such tests. All tests are solved using the implementation of the CRMHD equations as described in \Cref{subsec:numerical_methods} without stellar feedback and radiative cooling terms.

\subsubsection{Sod Shock Tube - Constant Efficiency}  \label{subsec:sodtest}

To test the validity of both our shock detection and CR injection schemes, we perform a Sod shock tube test, a standardised test used to benchmark the performance of the solver \citep{Sod1978}. We consider a 2D domain of $(x, y) \in [0, 10] \times [0,5]$ with an initial discontinuity at $x_0 = 5$. We set the initial densities on the left (post-shock, 1) and right (pre-shock, 2) sides of the discontinuities to be $\rho_1 = 1.0$ and $\rho_2 = 0.125$. The velocities on both sides of the discontinuity are set to zero, i.e. $u_1 =  u_2 = 0$. We choose the initial pressures to be $P_1 = 63.499$ and $P_2 = 0.1$ such that we obtain a Mach number $\mathcal{M}_1 = 10$ for a pure thermal gas and a Mach number of $\mathcal{M}_1 = 9.56$ for a composite thermal + CR gas. In this way, any deviations of the shock parameters are indications of the effect of the CR acceleration. We choose a thermal adiabatic index $\gamma_\mathrm{th} = 5/3$ and a CR adiabatic index $\gamma_\mathrm{CR} = 4/3$. To emphasise the effect of CR acceleration, we first take a constant acceleration efficiency $\eta = 0.5$. Here, we neglect magnetic fields (and in particular the dependence of the CR acceleration on $\theta_B$), CR diffusion, and hadronic losses to focus on the accuracy of our implementation. \par 

\begin{figure}[!h]
    \centering
    \resizebox{\hsize}{!}{\includegraphics{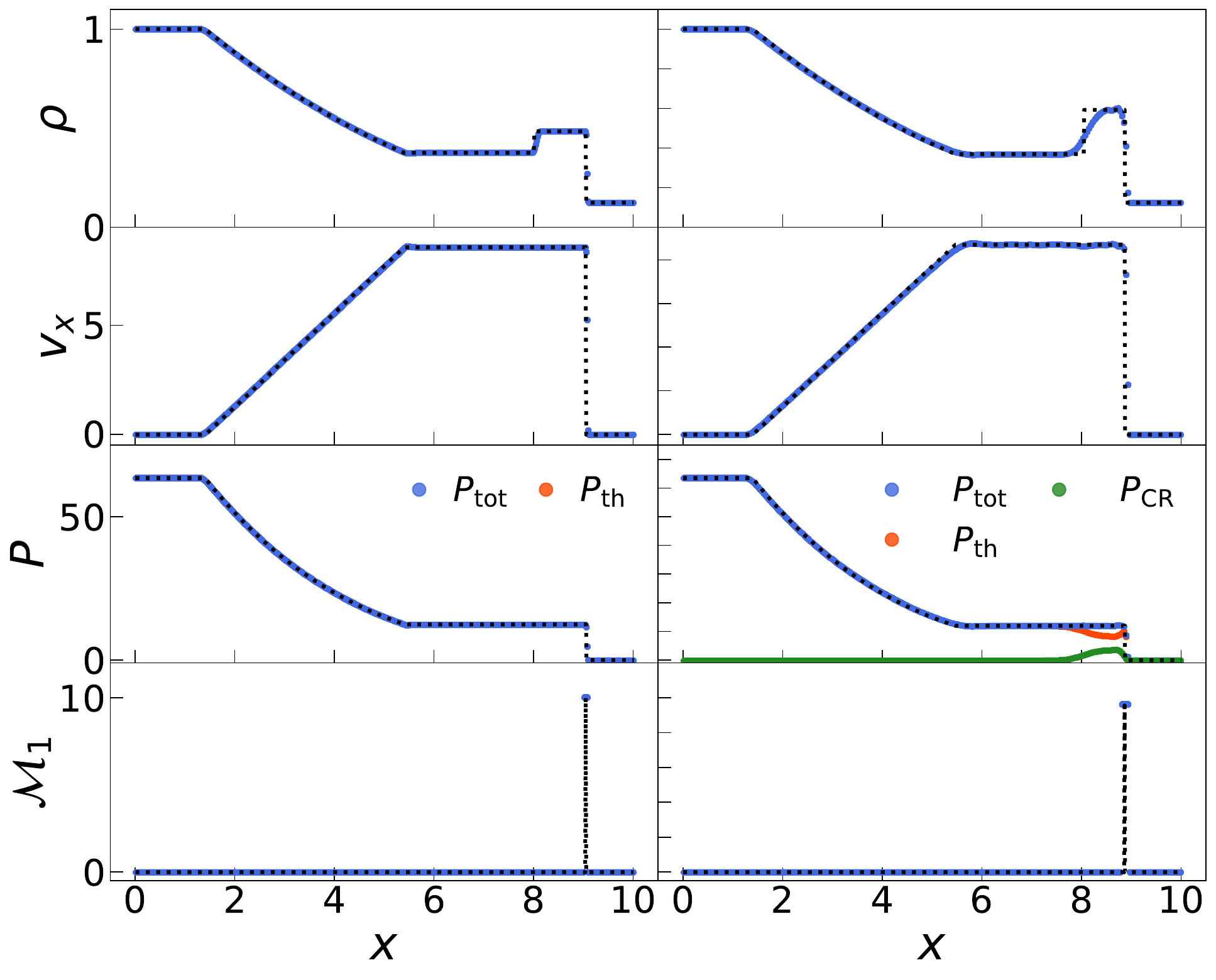}}
    \caption{Results from the 2D Sod shock tube test without (left) and with (right) CR injection at $t = 0.35$. We show the density, velocity, pressure, and the upstream Mach number averaged over the $y$-direction for both cases. We also plot the individual contributions of the pressure  ($P_\mathrm{th}$ without CR injection, $P_\mathrm{th}$ and $P_\mathrm{CR}$ with CR injection) for the pressure distribution. The analytical solution obtained from \cite{Sod1978} and \cite{Pfrommer2017} for the case without and with CR acceleration (black dotted line) are also shown.}
    \label{fig:CRinj_plot}
\end{figure}

\Cref{fig:CRinj_plot} shows the density, velocity, and pressure distribution averaged over the $y$-direction at $t = 0.35$ for the cases with and without CR injection. We perform our simulations with a uniform grid with a resolution of $\Delta x = 0.039$. We also overlay the analytical solution for both cases (without and with CR injection) obtained from \cite{Sod1978} and \cite{Pfrommer2017}, respectively. Overall, we observe excellent agreement between the numerical and analytical results for both cases. In particular, the shock is clearly captured in both cases as observed from the Mach number distribution, where the width of the `shock zone' is approximately 6 cells wide. However, the contact discontinuity is not properly captured in the case with CR injection, which is not a result of the CR injection scheme but rather due to numerical diffusivity of the solver \citep{Bouchut2010}. We further notice that the calculation of the Mach number in both cases shows excellent agreement with the analytical results, which we also show in a separate test to verify the calculation of the Mach number and dissipation energy (see \Cref{app:sod_mach}).

\subsubsection{Sod Shock Tube - Effect of Magnetic Obliquity}

We now test the validity of our implementation of the magnetic obliquity by performing a series of Sod shock tube tests but with varying magnetic obliquities. We use the same simulation setup as in \Cref{subsec:sodtest}.  However, to introduce the effect of the magnetic obliquity to the CR injection, we apply a magnetic field strength of $\vec{B} = (\num{1e-10}, \num{5.77e-11})$, $(\num{1e-10}, \num{1e-10})$, and $(\num{5.77e-11}, \num{1e-10})$ to generate magnetic obliquities of $\theta_B = 30^\circ, \, 45^\circ, \, $ and $60^\circ$, respectively, since the shock direction is directed only in the $\hat{x}$ direction. The magnetic field strengths are taken such that the magnetic pressure does not impact the overall dynamics of the simulation (i.e. $|\vec{B}|^2 \ll P$). In this limit, we can apply the analytical solution from \cite{Pfrommer2017} to verify the accuracy of our implementation. \par 

\begin{figure}[!h]
    \centering
    \resizebox{\hsize}{!}{\includegraphics{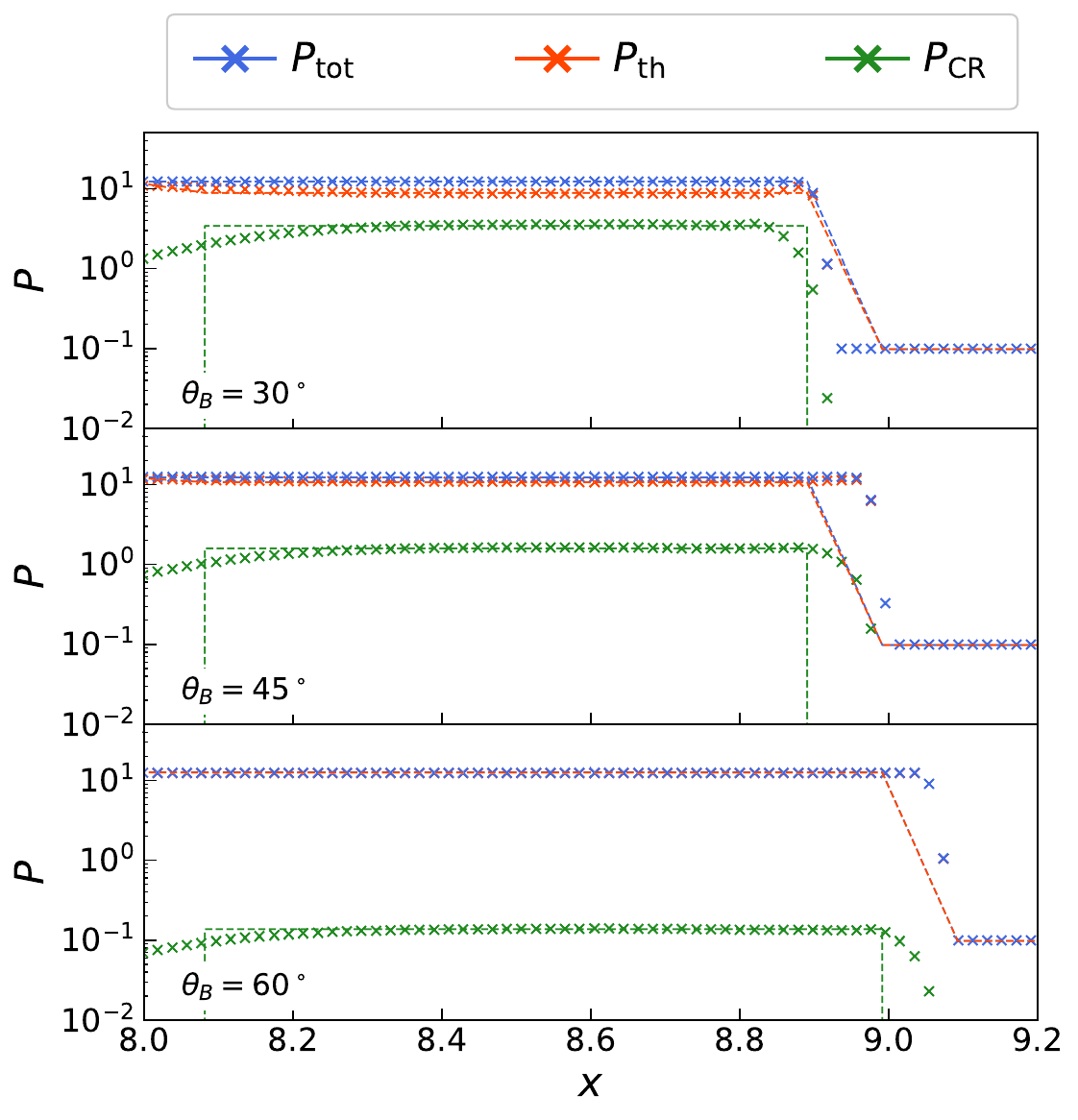}}
    \caption{Pressure distribution of the total pressure $P_\mathrm{tot}$ (blue), thermal pressure $P_\mathrm{th}$ (red), and CR pressure $P_\mathrm{CR}$ (green) from the 2-D Sod shock tube test with magnetic obliquities of $\theta_B = 30^\circ, \, 45^\circ, \, $ and $60^\circ$ at $t = 0.35$. The distribution is zoomed-in to the shocked region to better visualise the CR pressure at each obliquity. The dashed lines indicate the respective analytical solution for each pressure contribution as provided in \cite{Pfrommer2017}. The magnetic pressure contribution is not shown as they are negligible compared to the other pressure terms.}
    \label{fig:mgob_pres_plot}
\end{figure}

\Cref{fig:mgob_pres_plot} shows the pressure distribution of each component (total pressure $P_\mathrm{tot}$, thermal pressure $P_\mathrm{th}$ , and CR pressure $P_\mathrm{CR}$) for magnetic obliquities of $\theta_B = 30^\circ, \, 45^\circ, \, $ and $60^\circ$ at $t = 0.35$. We observe that for each obliquity we are able to reproduce the analytical solution from \cite{Pfrommer2017}, indicating excellent agreement at any value of magnetic obliquity. 

\subsubsection{Sedov-Taylor blast wave}  \label{subsec:sedovtest}

The Sedov-Taylor blast wave problem is a commonly used hydrodynamical test to analyse the performance of the solver with non-grid-aligned shocks. The problem solves a self-similar solution of a spherically expanding blast wave from a point explosion in a homogeneous medium \citep{Sedov1946}. We utilise this test to verify our shock detection implementation and in particular the capability to detect shocks in realistic scenarios. \par 

In our test, we consider a domain of $(x, y, z) \in [0, 1]^3 $ with an initial explosion energy $E_\mathrm{exp} = 1$ within an ambient density of $\rho_0 = 1$. We also set a small ambient pressure of $P_0 = \num{1e-5}$ everywhere except in the centre of the grid, and set an adiabatic index of $\gamma_\mathrm{th} = 1.4$. We do not consider CRs or magnetic fields in this test. \par 

We perform tests on an AMR grid with refinement levels determined by the grid size and spacing using the following prescription:
\begin{equation}
    l_\mathrm{AMR} = \left\lceil 1 + \log_2\left(\frac{x_\mathrm{max} - x_\mathrm{min}}{8\Delta x}\right) \right\rceil ,
    \label{eq:refinement_lvl}
\end{equation}
where $x_\mathrm{max}$ and $x_\mathrm{min}$ define the extent of the simulation domain. We start with a minimum refinement level of 0 with $\Delta x = 0.125$, and adaptively increase the refinement level by 1 after each timestep. We perform tests for two maximum refinement levels of 4 and 5, corresponding to resolutions of $\Delta x = 0.0078$ and $\Delta x = 0.0039$, respectively. This increases the number of cells per dimension from 8 (for level 0) to 256 (for level 4) and 512 (for level 5) respectively. \par 

\begin{figure}[!h]
    \centering
    \resizebox{\hsize}{!}{\includegraphics{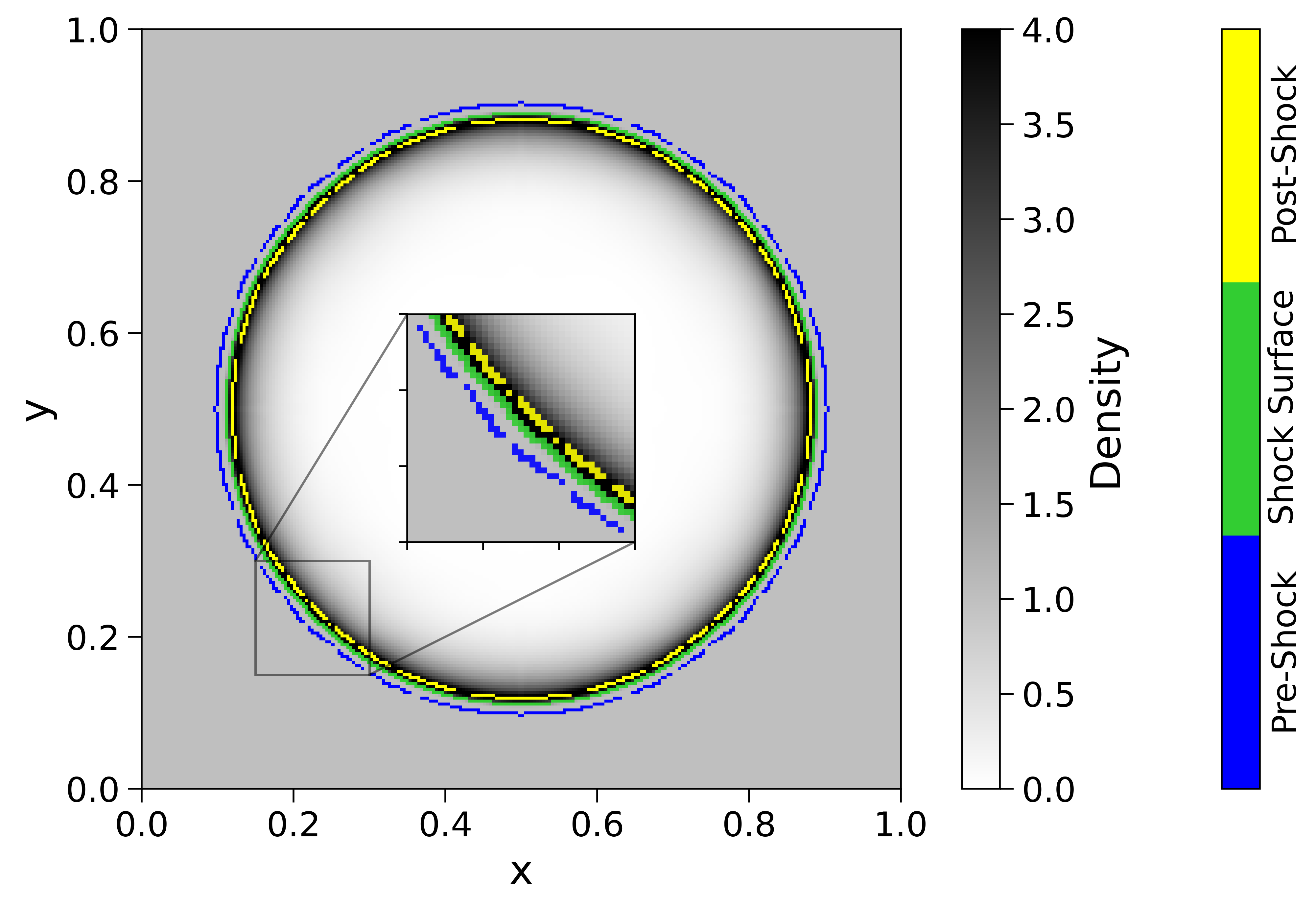}}
    \caption{Slice of the density distribution of the 3D Sedov test at $t = 0.09$ along $z = 0$. Here, a maximum refinement level of 4 is used, corresponding to 256 cells per dimension. We highlight the pre-shock, shock surface, and post-shock cells that are stored in the output files as blue, green, and yellow, respectively. We also show an inset plot zoomed in at $(x, y) \in [0.15, 0.3]^2$ to highlight the performance of the shock detection algorithm. The missing pre- and post-shock cells are contained in the guard cells, which are not stored in the snapshots. }
    \label{fig:sedov_dens}
\end{figure}

\Cref{fig:sedov_dens} shows a slice of the density distribution of the Sedov solution at $t = 0.09$ at $z = 0$. The pre-shocked cell, shock surface and post-shocked cell (blue, green and yellow respectively) determined from our implementation are also overlaid in the figure. \par

We observe that the shocked cells are captured well throughout the domain. The missing pre- and post-shock cells observed in the figure are contained in the guard cells, which are only stored locally per block and are not saved externally at each snapshot. As such, our shock detection algorithm works as intended, and this information is adequately propagated for the calculation of the relevant shock parameters as this computation is also performed locally within each block. The inset plot shows that the typical width of the `shock zone' is approximately 6 cells wide, comparable to those from the Sod tests. This indicates that realistic simulations that have non-grid-aligned shocks can also capture the shocked region with good agreement. We also see that the pre- and post-shock region is 2-3 cells away from the shock surface due to numerical broadening. \par 

To more quantitatively benchmark our algorithm, we compare the shock surface determined from our implementation with the analytical shock radius given by 
\begin{equation}
    R_\mathrm{sedov}(t) = \beta(\gamma_\mathrm{th}) \left(\frac{E_\mathrm{exp} t^2}{\rho_0}\right)^{1/5} = 0.394  \left(\frac{t}{0.09}\right)^{2/5},
    \label{eq:sedov_radius}
\end{equation}
where $\beta(\gamma_\mathrm{th} = 1.4) = 1.033$ is a dimensionless constant \citep{Landau1959}. \par

\begin{figure}[!h]
    \centering
    \resizebox{\hsize}{!}{\includegraphics{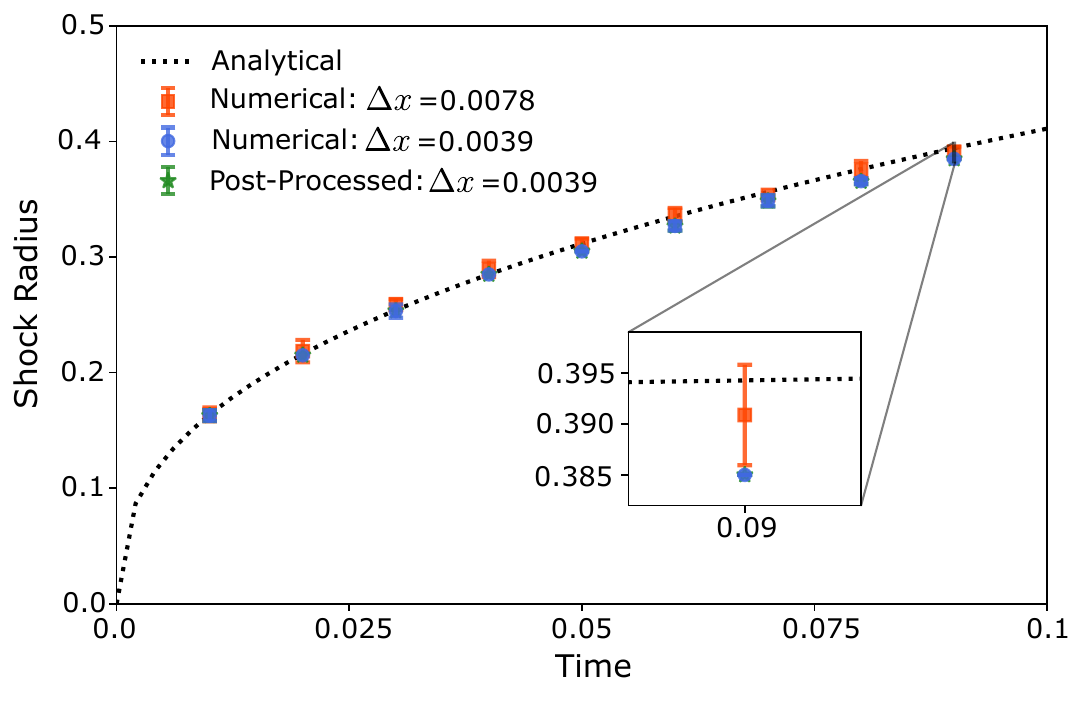}}
    \caption{Time evolution of the radius of the Sedov blast wave at maximum refinement level 4 and 5, corresponding to maximum resolutions of $\Delta x = 0.0078$ and $\Delta x = 0.0039$ (in orange and blue), respectively. We also show the post-processed results (in green) for the test with maximum refinement level 5. The post-processed results is directly above the numerical results from. The analytical Sedov radius from \Cref{eq:sedov_radius} is also shown. The inset plot shows the numerical and analytical values at $t = 0.09$ obtained from \Cref{fig:sedov_dens}. }
    \label{fig:sedov_rad_tevol}
\end{figure}

The shock radius is determined by taking the mean of the azimuthal average of the determined shock surface. In \Cref{fig:sedov_rad_tevol}, we show the time evolution of the shock radius determined with a maximum refinement level of 4 and 5 in the Sedov test, corresponding to a maximum resolution of $\Delta x = 0.0078$ and $\Delta x = 0.0039$, respectively. We also compare our on-the-fly implementation with an identical implementation performed in post-processing. Here, we took the results for the simulation with a maximum refinement level of 5 at each snapshot and applied the same shock detection implementation to identify the shocked regions. Overall, our implementation shows excellent agreement with the post-processed results, which can be observed from the negligible error bars between the two results. Furthermore, the scaling relation of our implementation is consistent with the analytical results except from a systematic deviation of $\sim 2\%$ at later times, primarily attributed to the numerical diffusivity of the MHD solver and the numerical broadening of the shocked region within our implementation. \par  

    \section{CR acceleration from runaway stars}
    \label{sec:runawaystars}
    We now apply our shock detection and CR injection implementation to simulate the CR acceleration in bow shocks forming around massive runaway stars and study the impact on their morphology and evolution. 

\subsection{Simulation setup}   \label{subsec:sim_setup}

\begin{table*}
    \caption{Models used to simulate stellar bow shocks in our work. The star velocities and diffusion coefficients are modified to simulate runaway stars in different scenarios. The characteristic time scales and length scales set for each model, as well as the maximal spatial resolution, $\Delta x$, used for each simulation, are also shown. }
    \label{tab:sim_setup_vals}
    \centering
    \begin{tabular}{ccccccccc}
    \hline \hline
        Model & $v_\star \, /\, \si{\kilo\meter\per\second}$ & $\kappa_{\parallel, 10} \, / \, \si{\centi\meter\squared\per\second}$ & $\Delta x$ / $\si{\parsec}$ & $R_0$ / $\si{\parsec}$ & $r_\mathrm{w,inj}$ / $\si{\parsec}$ &  $t_\mathrm{dyn}$ / $\si{\kilo\year}$ & $t_\mathrm{CR, diff}$ / $\si{\kilo\year}$ & Comments \\ 
        \hline
        \texttt{ISM-v40-LDiff} & 40 & $\num{3e24}$ & 0.195 &  1.02 & 0.39  & 25.1 & 31.8 & \\
        \texttt{ISM-v40-MDiff} & 40 & $\num{3e25}$ & 0.195 & 1.02 & 0.39 & 25.1 & 3.2 &  Reference Model \\
        \texttt{ISM-v40-HDiff} & 40 & $\num{3e26}$ & 0.195& 1.02 & 0.39 & 25.1 & 0.3 & \\
        \texttt{ISM-v100-MDiff} & 100 & $\num{3e25}$ & 0.098 & 0.43 & 0.20 & 4.2 & 5.6 &  \\
        \texttt{ISM-v40-MHD} & 40 & ... & 0.195 & 1.32 & 0.39 & 25.1 & ... &  No CR injection \\ 
        \hline
    \end{tabular}
\end{table*}

In this work, we consider a 3D AMR grid in Cartesian geometry with a spatial domain of $(x, y, z) \in \left[-50, 50\right] \times \left[-25, 25\right] \times \left[-25, 25\right]$ pc. We perform the simulation in the stellar rest frame throughout the simulation, where $+\hat{x}$ points along the motion of the star (i.e. bow axis) and $\hat{y}$ and $\hat{z}$ are the perpendicular directions. The simulation domain here is extended only in the $x$ dimension to ensure that the bow shock is fully captured throughout the simulation. We place a sink particle (representing the star) at the origin stationary at all times and set the velocity of the ambient gas directly opposite that of the propagation velocity of the star, i.e. $\vec{u} = -v_\star \hat{x}$. To model this throughout the domain, we apply inflow and outflow boundary conditions on the left and right boundary along the $x$-direction while keeping periodic boundary conditions in the $y$- and $z$-directions. \par 

In all models, the star has a mass of $M_\star = \SI{20}{\solarmass}$, and to investigate the impact of CR shock acceleration on the early evolution of the wind-driven bow shock, we run the simulations until $t = \SI{500}{\kilo\year}$, where the star velocity $v_\star$ is greater than the expansion of the wind bubble $R_\mathrm{wind}(t) / t$ \citep{Weaver1977}. To describe the mass loss rate and wind velocity of the star, we follow the stellar evolution tracks for a zero-age main sequence star from \citet{Ekstroem2012}. Due to the short timescale of the simulation, both values are approximately constant in time, with a time-averaged value of $\dot{M} \sim \SI{2.6e-8}{\solarmass\per\year}$ and $v_\mathrm{wind} \sim \SI{2.7e3}{\kilo\meter\per\second}$, respectively. This yields a wind luminosity of $L_w \sim \SI{6e34}{\erg\per\second}$ throughout the simulation, which is used to calculate the injected momentum from stellar winds, $\dot{\vec{q}}_\mathrm{wind}$, at each timestep (see \Cref{sec:numerical_impl} and \citealt{Gatto2017} for more details). We do not include stellar magnetic fields in the setup. \par 

For all simulations, we emulate the surrounding medium as an ISM-like environment by choosing an ambient density of $\rho_0 = \SI{2e-25}{\gram\per\centi\meter\cubed}$, ambient gas temperature $T_0 = \SI{1e4}{\kelvin}$, and a weak uniform magnetic field strength of $\vec{B} = B \; \hat{x} = \SI{1e-6}{\gauss} \; \hat{x} $ parallel to the direction of propagation. We also perform simulations for stellar velocities of $40$ and $\SI{100}{\kilo\meter\per\second}$ \citep[motivated from stellar velocity distributions from][]{Renzo2019}, which we label as \texttt{ISM-v40-MDiff} and \texttt{ISM-v100-MDiff} models, respectively. As the Galactic (spatial) CR diffusion coefficient is not well constrained, we also explore the impact of different diffusion coefficients on the dynamics of the bow shock. We modify the parallel component of the diffusion coefficient at $E_\mathrm{CR} = \SI{10}{\giga\electronvolt}$ ($\kappa_\mathrm{\parallel, 10}$) and set the perpendicular component to 0.01$\kappa_{\parallel, 10}$ \citep{Nava2013}. We choose values of $\SI{3e24}{\centi\meter\per\second\squared}$ and $\SI{3e26}{\centi\meter\per\second\squared}$ (\texttt{ISM-v40-LDiff} and \texttt{ISM-v40-HDiff}, respectively) to investigate the effect of lower and higher diffusion coefficients compared to those of \texttt{ISM-v40-MDiff} (with $\kappa_{\parallel, 10} = \SI{3e25}{\centi\meter\per\second\squared}$). We also perform a simulation without any CR injection (\texttt{ISM-v40-MHD}) to highlight the impact of CR acceleration on the evolution of the bow shock. \par

The characteristic length scale of the system is defined through the stand-off radius, $R_0$, evaluated from the pressure balance between the wind and the ISM:
\begin{equation}
    R_0^2 = \frac{\dot{M} v_\mathrm{wind}}{4\pi \rho_0\left(v_\star^2 + c_\mathrm{ISM}^2 + v_A^2\right)},
    \label{eq:standoff_radius}
\end{equation}
where $c_\mathrm{ISM}$ and $v_A$ are the isothermal sound speed and Alfv\'{e}n speed of the surrounding medium, respectively. The stand-off radius, in the limit where cooling is instantaneous, corresponds to the distance from the star to the apsis of the bow shock. \par 

As all parameter values in \Cref{eq:standoff_radius} are known in the simulation setup, we use the stand-off radius to set the maximal refinement level (as defined in \Cref{eq:refinement_lvl}) necessary for each simulation. This is set by ensuring that the wind injection radius, $r_\mathrm{w,inj} = 4 \times \Delta x$, is at least smaller than the stand-off radius, i.e. $r_\mathrm{w,inj} < R_0$. From this criterion, we choose a maximal refinement level of 6 and 7 (with corresponding maximal spatial resolution of $\Delta x = \SI{0.195}{\parsec}$ and $\SI{0.097}{\parsec}$) for simulations with $v_\star = \SI{40}{\kilo\meter\per\second}$ and $v_\star = \SI{100}{\kilo\meter\per\second}$, respectively. This sets the maximal number of cells per dimension to $512 \times 256 \times 256$ and $1024 \times 512 \times 512$, respectively. Through the stand-off radius, we can also compute the dynamical timescale, $t_\mathrm{dyn} = R_0 / v_\star$, and the CR diffusion timescale, $t_\mathrm{CR, diff} = R_0^2 / \kappa_{\parallel, 10}$, for each model. \Cref{tab:sim_setup_vals} shows each model used in this work, listing the characteristic timescale and length scales for each model. \par

\subsection{Dynamical Impact of CR shock acceleration on Magnetohydrodynamical Parameters}

\begin{figure*}
    \centering
    \includegraphics[width=16cm]{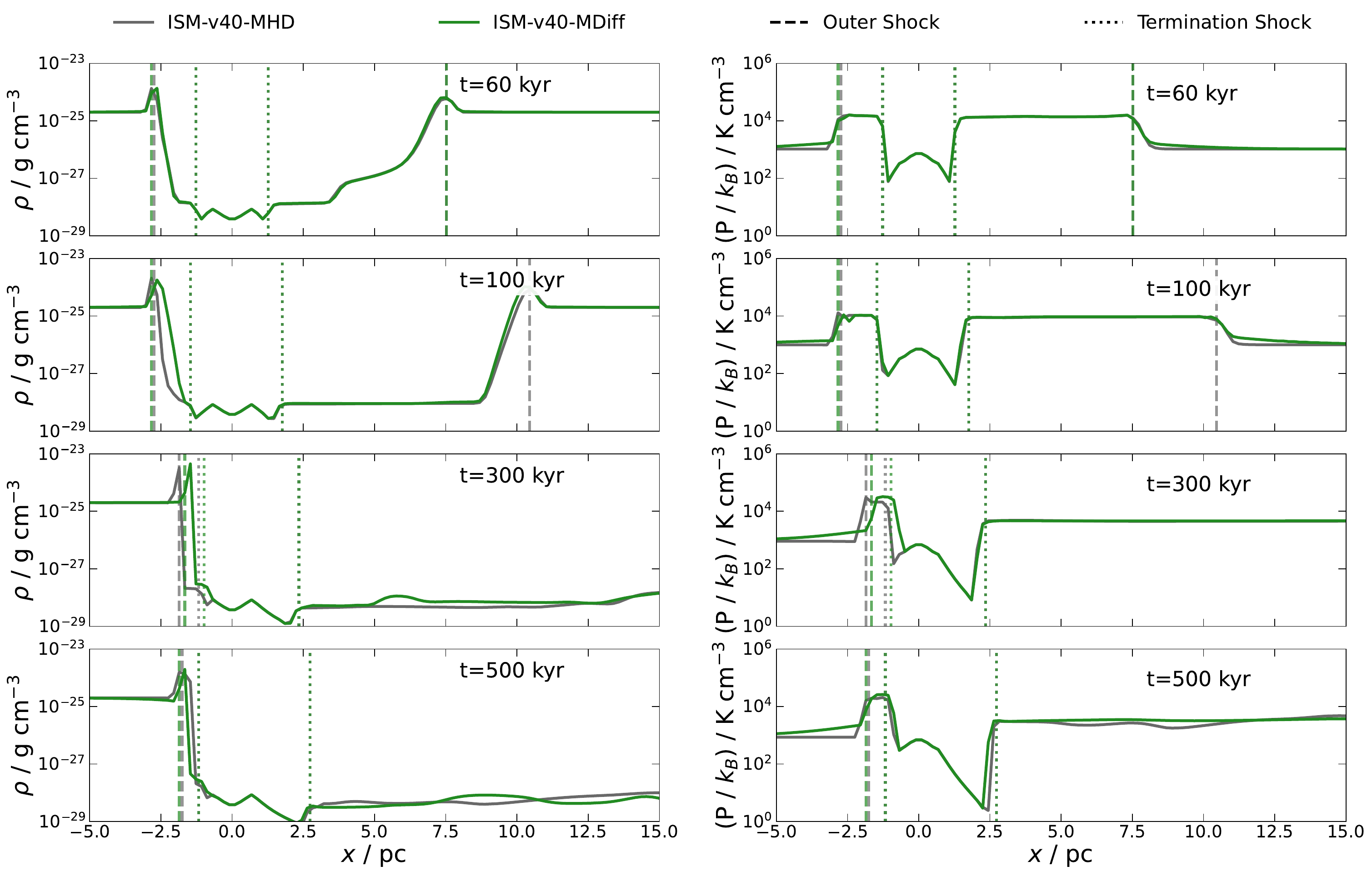}
    \caption{1D slices of the density (left) and total pressure (right) along the bow axis (i.e. at $y = z = 0$) for the \texttt{ISM-v40-MHD} (gray) and \texttt{ISM-v40-MDiff} (green) models at (from top to bottom) $t = 60, 100, 300$, and $\SI{500}{\kilo\year}$. Here the outer (dashed) and termination shock (dotted) as identified by our shock detection implementation is highlighted for each timestep. $-\hat{x}$ points towards the apsis of the bow shock, and the star is located at $x=0$. }
    \label{fig:crmhd_vs_mhd_1dslice}
\end{figure*}

In \Cref{fig:crmhd_vs_mhd_1dslice}, we show the 1D slice of the density and pressure along the $x$-axis, while taking $y = z = 0$. We overlay results from the reference case of \texttt{ISM-v40-MDiff} and \texttt{ISM-v40-MHD} models at $t = 60, 100, 300$, and $\SI{500}{\kilo\year}$.
We also indicate the outer and termination shocks in each subfigure of \Cref{fig:crmhd_vs_mhd_1dslice}. As the shock-finding algorithm cannot distinguish between the outer and termination shocks, we filter the shocked cells corresponding to the termination shock by considering cells that have densities $\rho \leq \SI{1e-27}{\gram\per\centi\meter\cubed}$ as motivated from typical densities near the free wind region, and take all shocked cells outside this criterion as those originating from the outer shock. We also apply a pressure threshold value of $\num{4e5} k_B^{-1} \si{\kelvin\per\centi\meter\cubed}$ to filter out any additional numerical artefacts detected by our shock algorithm\footnote{Due to the numerical broadening of the shock zone within our implementation, we consider post-shock states by taking quantities masked by both post-shock and shock surface cells as defined within our algorithm (see \Cref{subsec:shock_det}).}. \par 

We observe that the morphology of the density profile for both models are similar at all times, except at the regions near the forward shock. In particular, while the MHD results approach towards a steady-state solution, as indicated by the lack of variation of the apsis position, the CRs that diffuse outward from their injection sites induces an additional pressure term that `smears' out the forward shock. This CR pressure act onto the apsis of the bow shock, delaying the rate in which the bow shock achieves steady-state. This is clear by the non-negligible pressure contribution directly outside the apsis of the shock, which is not observed in the MHD case. \par

\begin{figure*}
    \centering
    \includegraphics[width=15cm]{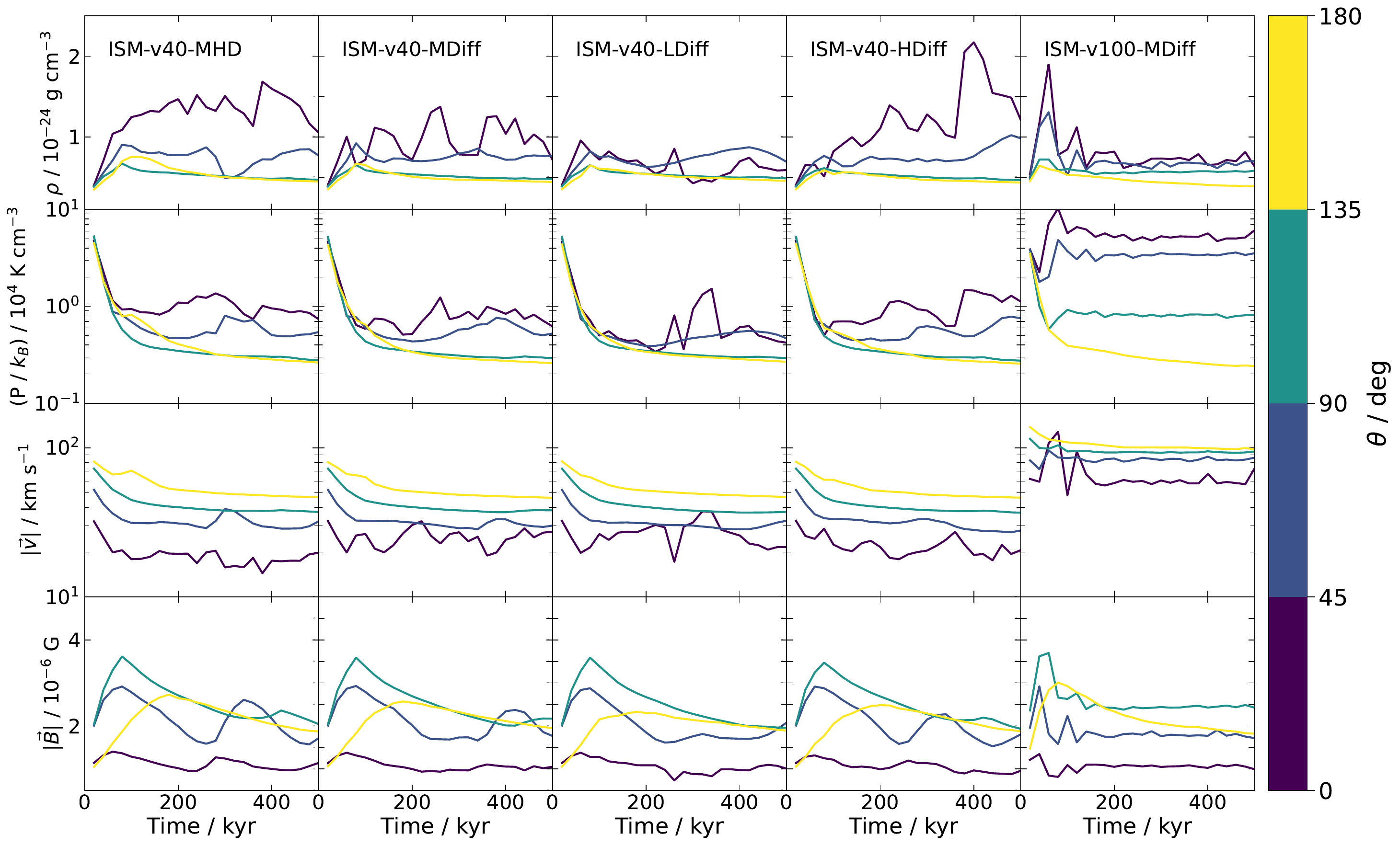}
    \caption{Time evolution of the (from top to bottom) density, magnitude of the velocity, pressure, and magnetic field strength for all models at the outer shock considered in this work. The colorbar indicates the dependence on the polar angle, where $\theta = \SI{0}{\degree}$ points towards the apsis of the bow shock. The time evolution of each parameter is obtained by taking the azimuthal and radial average from those at the outer shock (identified by excluding all cells identified via the shock detection algorithm with densities $\rho \leq \SI{1e-27}{\gram\per\centi\meter\cubed}$).}
    \label{fig:crmhd_vs_mhd_polar_time}
\end{figure*} 

\begin{figure*}[!h]
    \centering
    \includegraphics[width=17cm]{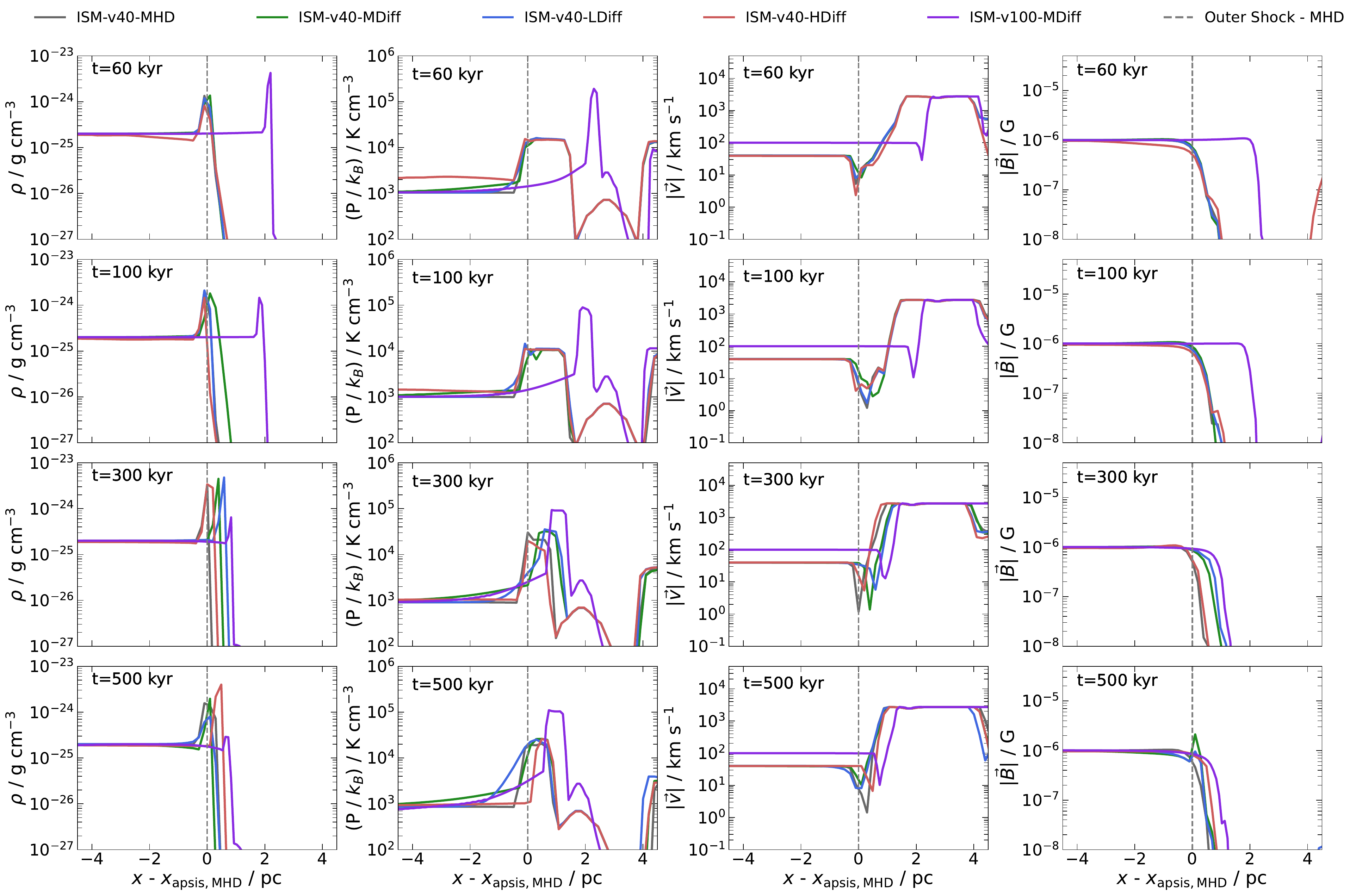}
    \caption{1D slices of the shock front around the apsis of the bow shock for the (from left to right) density, pressure, magnitude of velocity and magnetic field strength at (from top to bottom) $t = 60, 100, 300$, and $\SI{500}{\kilo\year}$. The slice is taken at $y = z = 0$ for all models, and the $x$-axis is shown relative to the position of the apsis of the bow shock from the \texttt{ISM-v40-MHD} model, $x_\mathrm{apsis, MHD}$, at all times. }
    \label{fig:crmhd_vs_mhd_1d_nearshock}
\end{figure*} 

We now compare the results from the \texttt{ISM-v40-MHD} case with all other models considered in this work. \Cref{fig:crmhd_vs_mhd_polar_time} shows the time evolution of the density, pressure, magnitude of the velocity, and magnetic field strength for all models at the post-shock states of the outer shock. The dependence on polar angle, $\theta$ (where $\theta = \SI{0}{\degree}$ points toward the apsis of the bow shock), is calculated by taking the azimuthal and radial average of the CR shock acceleration parameters at the shock as identified using the shock-finding algorithm as described in \Cref{subsec:shock_det}. The outer shock is determined using the selection criterion as described above. The average value of each parameter is then taken for each range of polar angle. In this way, we showcase the spatial dependence along the bow shock. \par 

\begin{figure*}[t]
    \centering
    \includegraphics[width=17cm]{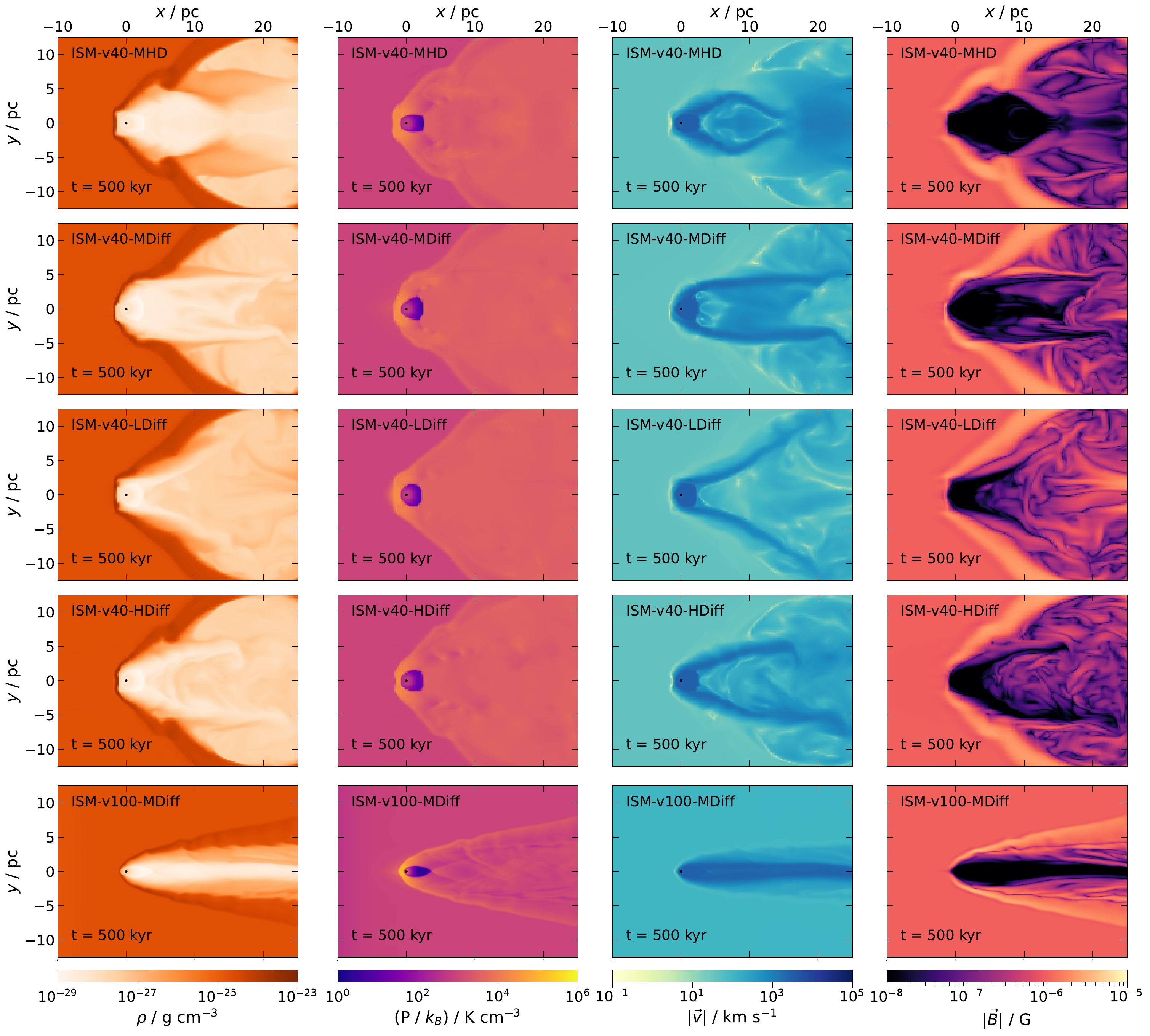}
    \caption{2D slices of the (from left to right) density, pressure, magnitude of the velocity, and magnetic field strength distribution of the (from top to bottom) \texttt{ISM-v40-MHD}, \texttt{ISM-v40-MDiff}, \texttt{ISM-v40-LDiff}, \texttt{ISM-v40-HDiff}, and \texttt{ISM-v100-MDiff} models at $t = \SI{500}{\kilo\year}$. The slices are taken at $z = 0$ as the bow shock is axisymmetric about the $x$-axis.}
    \label{fig:crmhd_vs_mhd_2dslice}
\end{figure*}

To further highlight the morphological differences between different models, we also show the time evolution of the 1D slice (at $y = z = 0$) of the same parameters around the apsis of the bow shock in \Cref{fig:crmhd_vs_mhd_1d_nearshock}. The figure is centred based on the apsis of the bow shock in the MHD case, determined by our shock detection algorithm, at each time. \Cref{fig:crmhd_vs_mhd_2dslice} shows the 2D slice of the same MHD parameters at $t = \SI{500}{\kilo\year}$. \par 

Comparing the \texttt{v40} models with the MHD case, we observe that, morphologically, the \texttt{HDiff} case is most similar to that of the MHD case. With a stronger CR diffusion strength, most of the CRs have been diffused out of the bubble, resulting in a reduced CR pressure that acts on the bow shock. In contrast, the stronger CR pressure term in the \texttt{LDiff} model greatly modifies the morphology of the bow shock. In all cases, we also observe that the winds within the bow shock are distorted from a perfectly axisymmetric shape, as observed in the MHD case. \par

When comparing the features near the apsis of the shock, we observe that all models approach the same apsis position as with the MHD case (again, with the exception of the \texttt{v100} case). Only minor differences in the density and velocity strength at the shock can be observed. This indicates that the inclusion of CR acceleration in bow shocks does not significantly modify features of the bow shock, but only morphological features as observed in \Cref{fig:crmhd_vs_mhd_2dslice}. \par 

The \texttt{ISM-v100-MDiff} case shows significant variation with other models. But this is expected, as the higher velocity results in a shorter dynamical timescale (as shown in \Cref{tab:sim_setup_vals}). Nevertheless, within the same timesacles, the position of the apsis approaches closer to that of the MHD case, even with a higher velocity. This shows that the bow shock from all other models has reached its steady-state. Only the pressure distribution in \Cref{fig:crmhd_vs_mhd_polar_time} shows significant deviation from the expected behaviour, which is likely due to the stronger kinetic pressure from the ambient medium.

\begin{figure*}[t]
    \centering
    \includegraphics[width=17cm]{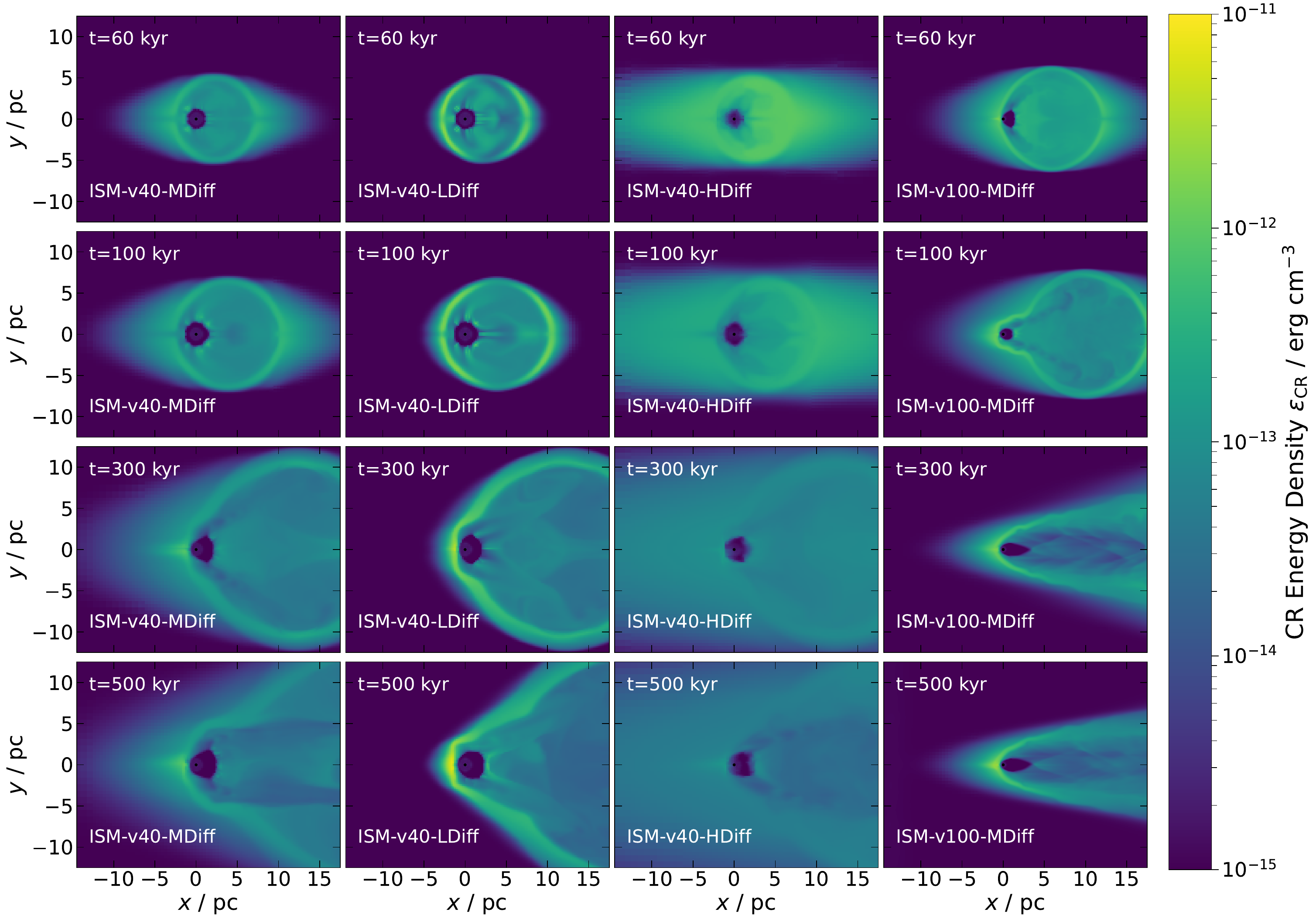} 
    \caption{Time evolution of the 2D slices of the CR energy density, $\epsilon_\mathrm{CR}$, for the (from left to right)\texttt{ISM-v40-MDiff}, \texttt{ISM-v40-LDiff}, \texttt{ISM-v40-HDiff}, and \texttt{ISM-v100-MDiff} models considered in this work at $t = 60, 100, 300$, and $\SI{500}{\kilo\year}$.}
    \label{fig:encr_timeevo_2d}
\end{figure*}

\subsection{Early evolution of CR parameters at the shock}  \label{subsec:shock_params}

\Cref{fig:encr_timeevo_2d} shows the time evolution of the 2D slice of the CR energy density for all models considered in this work. We already observe a strong polar angular dependence at earlier times. This is due to the obliquity-dependent CR acceleration efficiency as stated in \Cref{eq:craccel_mgob}, where parallel shocks are most efficient for CR injection. At later times, the expansion of the wind bubble reduces the total CR energy near the shock. \par 

It is also clear from the comparison of the CR energy density distributions that a higher diffusion coefficient allows CRs that are accelerated at the outer shock to propagate away from the wind bubble more rapidly, thus reducing the available CRs within the bubble. As such, the density distribution for \texttt{ISM-v40-HDiff} resembles that of the MHD case. In contrast, the \texttt{ISM-v40-LDiff} case show that the CRs are continuously injected at the shocks in the termination shock, allowing a higher rate of CR production, increasing the overall CR acceleration near the shocks. We also observe that the overall CR energy density varies with the polar angle with respect to the apsis of the bow shock, where more CRs are injected along the $x$-axis. The behaviour of the CRs between the \texttt{ISM-v40-MDiff} and \texttt{ISM-v100-MDiff} case are similar in both cases, except from kinematic differences due to the higher stellar velocity. 

\begin{figure*}
    \centering
    \includegraphics[width=15cm]{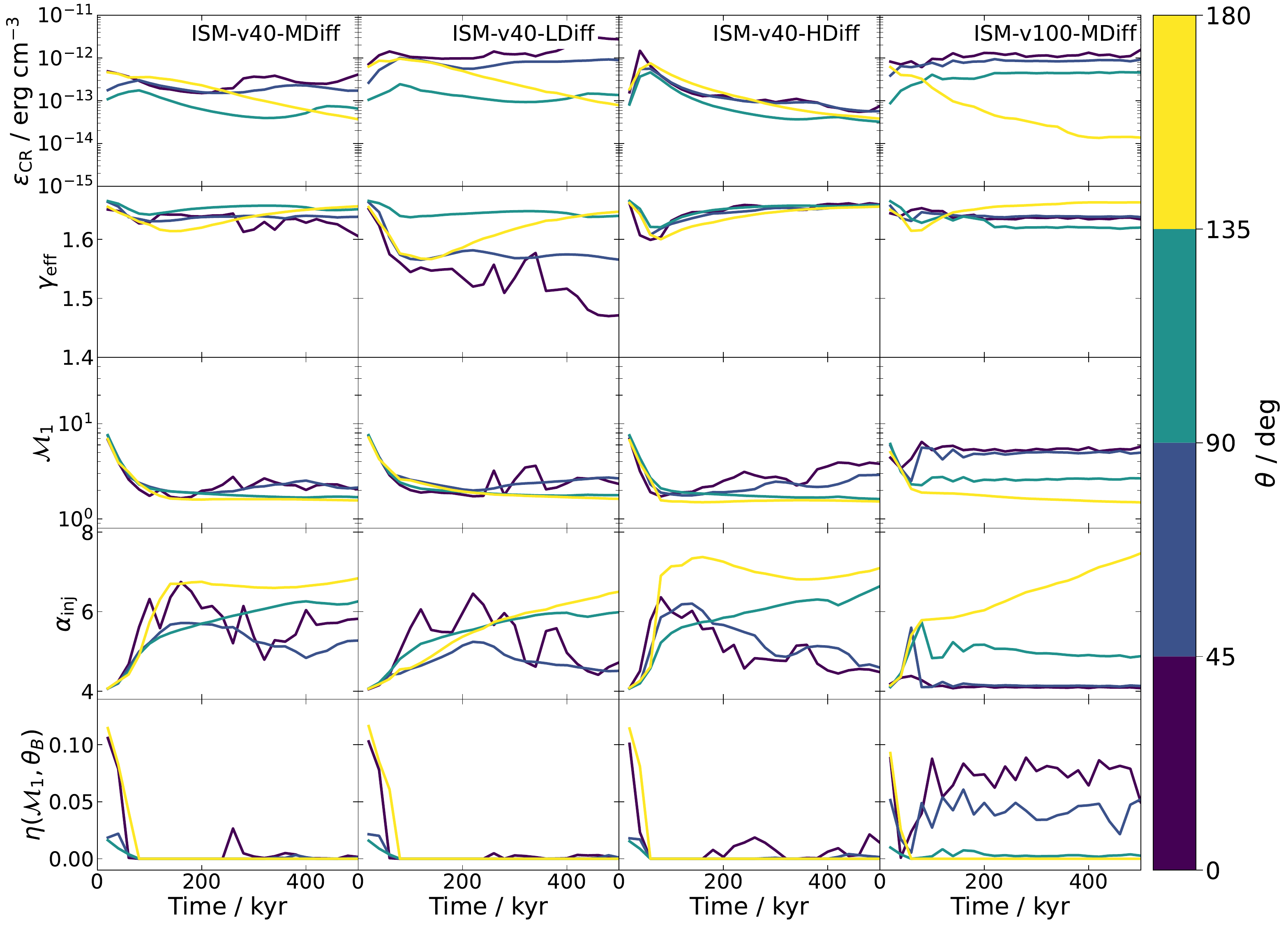} 
    \caption{Same as \Cref{fig:crmhd_vs_mhd_polar_time}, but for the (in order from top to bottom) CR energy density, $\epsilon_\mathrm{CR}$, effective adiabatic index, $\gamma_\mathrm{eff}$, pre-shock Mach number, $\mathcal{M}_1$, injection index, $\alpha_\mathrm{inj}$,  and CR acceleration efficiency, $\eta$.}
    \label{fig:crshockparams_outer}
\end{figure*}

\begin{figure*}
    \centering
    \includegraphics[width=15cm]{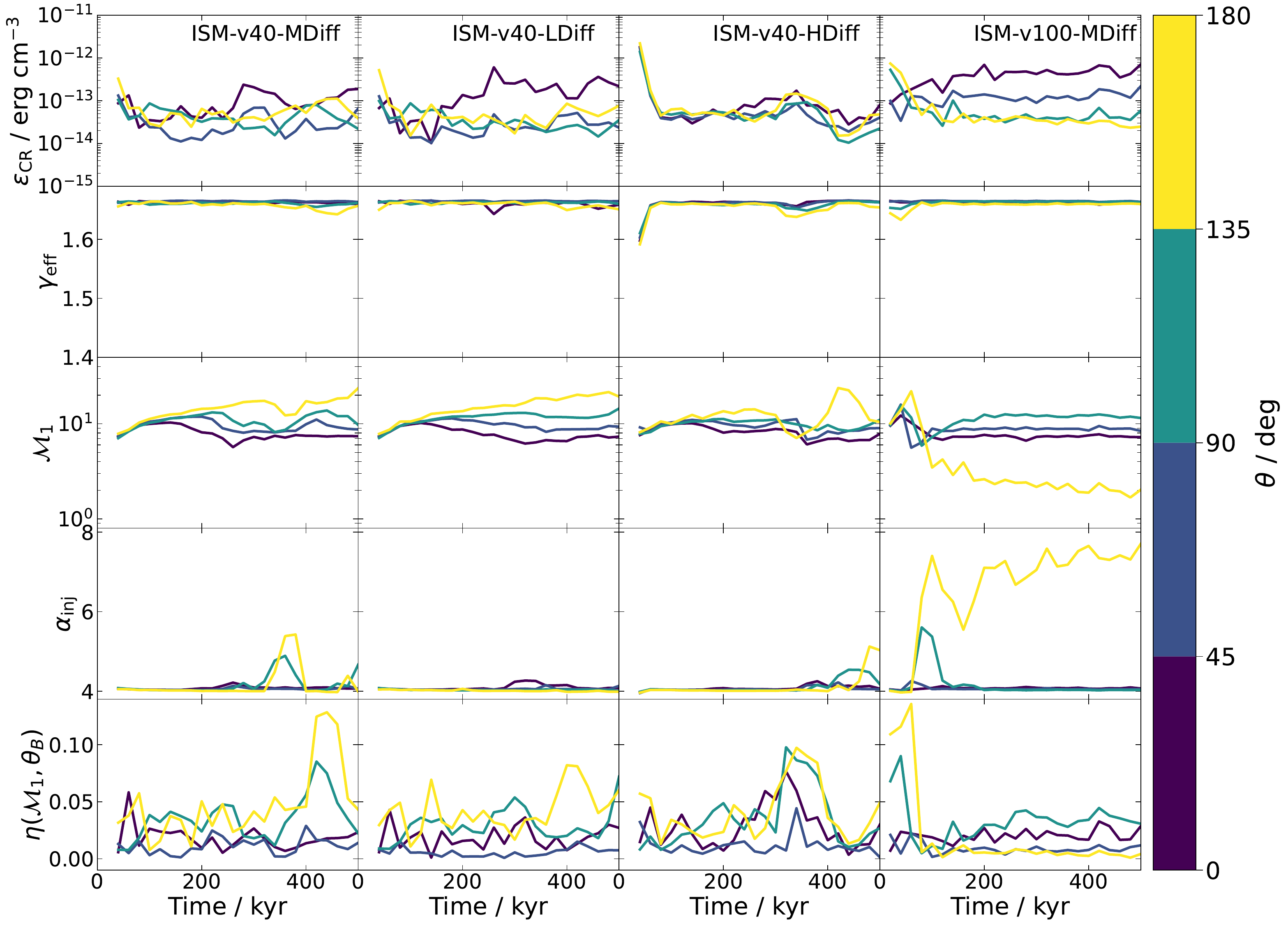} 
    \caption{Similar to \Cref{fig:crshockparams_outer} but taking the CR shock acceleration parameters at the termination shock instead, where the termination shock is determined by marking all identified cells via the shock detection algorithm with densities  $\rho \leq \SI{1e-27}{\gram\per\centi\meter\cubed}$.}
    \label{fig:crshockparams_term}
\end{figure*}

\Cref{fig:crshockparams_outer} and \Cref{fig:crshockparams_term} show the time evolution of the CR shock acceleration parameters, namely the CR energy density, $\epsilon_\mathrm{CR}$, the effective adiabatic index, $\gamma_\mathrm{eff}$, the pre-shock Mach number, $\mathcal{M}_1$, the CR acceleration efficiency, $\eta$, and the injection spectral index $\alpha_\mathrm{inj}$ for each model at the post-shock states observed at the outer shock and termination shock, respectively. The injection spectral index (in momentum space), $\alpha_\mathrm{inj} = \alpha_\mathrm{inj}(\mathcal{M}_1)$, is calculated following arguments from diffusive shock acceleration \citep{Vainio1999}: 
\begin{equation}
    \alpha_\mathrm{inj}(\mathcal{M}_1) = \frac{3 x_s(\mathcal{M}_1)}{x_s(\mathcal{M}_1) - 1}.
    \label{eq:inj_index}
\end{equation}
where $\alpha_\mathrm{inj} \rightarrow 4$ for strong shocks. \par 

Initially, we observe that the outer shock has Mach numbers of $ \sim 10$, with $\eta \sim 0.1$, which induces CR acceleration at the earliest times. However, the Mach number at the outer shock rapidly approaches to $O(1)$, while the Mach number of the termination shock increases to $> 10$. This allows for CRs to be accelerated only at the termination shock for $t \gtrsim \SI{50}{\kilo\year}$. This is also clear when looking at the spectral index; $\alpha_\mathrm{inj} \rightarrow 4$ for all times for the termination shock, while the values quickly increase due to the weak outer shock. This indicates that the termination shock is a constant acceleration site for CRs for bow shocks, consistent with e.g. \cite{BenRomMar10,DelRom12, DelBosMul18,BenDelHal21,Martinez:2023}. \par 

We also observe that even slow shocks with $\mathcal{M}  \lesssim 10$ can accelerate CRs. This is a limitation of our algorithm, which cannot distinguish between the acceleration efficiency of low-velocity and high-velocity shocks. This limited treatment of velocity-dependent CR acceleration can overestimate the overall efficiency of CR accelerations. Even with disallowing shocks with $\mathcal{M} \leq 3$ to accelerate CRs, with most of the shocks with $\mathcal{M} \sim 10$, the acceleration will nevertheless be overestimated (see \Cref{app:mach_eff}). As such, the results can only be considered as an upper estimate of the predicted acceleration efficiency. \par 

The CR energy and $\gamma_\mathrm{eff}$ do not vary strongly with the polar angle, except for those near the periapsis of the bow shock. This is likely due to the weaker shocks as the periapsis is the furthest from the stellar winds, which is the prominent source of shocks in this case. Only a mild dependence on the Mach number can be observed with the polar angle at the termination shock, as the `bell-like' structure, commonly observed in bow shocks, has fewer acceleration sites in the forward direction. \par  

Between the diffusion models, only the CR energy and $\gamma_\mathrm{eff}$ vary strongly, which is a direct consequence from the CRs diffusing away from the bubble (with a higher CR diffusion coefficient). All other acceleration parameters do not strongly vary between the diffusion models, indicating that the CR shock parameters are dominated by the stellar kinematics. The \texttt{v100} run, interestingly, shows that after $t \sim   \SI{50}{\kilo\year}$, the outer shock in the forward direction can further accelerate CRs, while CRs are accelerated less at the termination shock. Comparing the temporal behaviour of the \texttt{v100} run with the \texttt{v40} runs, we also observe that the behaviour of the \texttt{v40} run up to $t = \SI{500}{\kilo\year}$ is akin to that of $t \sim \SI{50}{\kilo\year}$ for the \texttt{v100} run. This indicates that, at later times, the outer shock can potentially be an acceleration site for CRs as well. This can be confirmed by running the other models for a longer runtime, which we leave for future studies.

    \section{Synthetic Observations}
    \label{sec:observations}
    In this section, we use our results from our simulations to estimate the population of CR protons and electrons within the medium. Using this information, we estimate the gamma-ray emission originating from pion decay and synchrotron emission resulting from CR electrons produced through hadronic collisions of CR protons, which we can directly compare with observed values from bow shocks within the ISM.

\subsection{Gamma-ray emission}  \label{subsec:gamma_ray}

Gamma-ray emission can be observed from bow shocks through the $\pi^0$-decay originating from hadronic collisions between CR protons and thermal protons within the environment. The gamma-ray luminosity can be estimated using similar arguments to describe the hadronic losses, however, taking into account that only $\approx 1/3$ of the collision processes yield neutral pions. Following \cite{Longair2011}, the gamma-ray luminosity can be calculated as such:
\begin{equation}
    L_\gamma = \int \diff^3 x j_\gamma(\vec{x}) =  \frac{1}{3} \frac{c \sigma_{pp} K_p}{\mu m_p} \: \int \diff ^3 x \: \rho(\vec{x}) \epsilon_\mathrm{CR}(\vec{x}),
    \label{eq:gamma_lum}
\end{equation}
where $\sigma_{pp} \approx \SI{44}{\barn} = \SI{4.4e-23}{\centi\meter\squared}$ is the total pion cross section \citep{Pfrommer2004} and $K_p \approx 1/2$ is the inelasticity of the interaction. The integration is performed over the density and CR energy density in each cell for the whole simulation domain, as the integral over CR energy is already implicitly performed in the CR transport equation. Due to the nature of our injection algorithm, without a spectrally resolved CR model, we cannot distinguish the fraction of CR protons that have sufficiently high enough energy to participate in p-p interactions. As such, we note that the results we show here should be considered as upper-limits of the gamma-ray emission.  \par 

\begin{figure*}[!h]
    \centering
    \includegraphics[width=17cm]{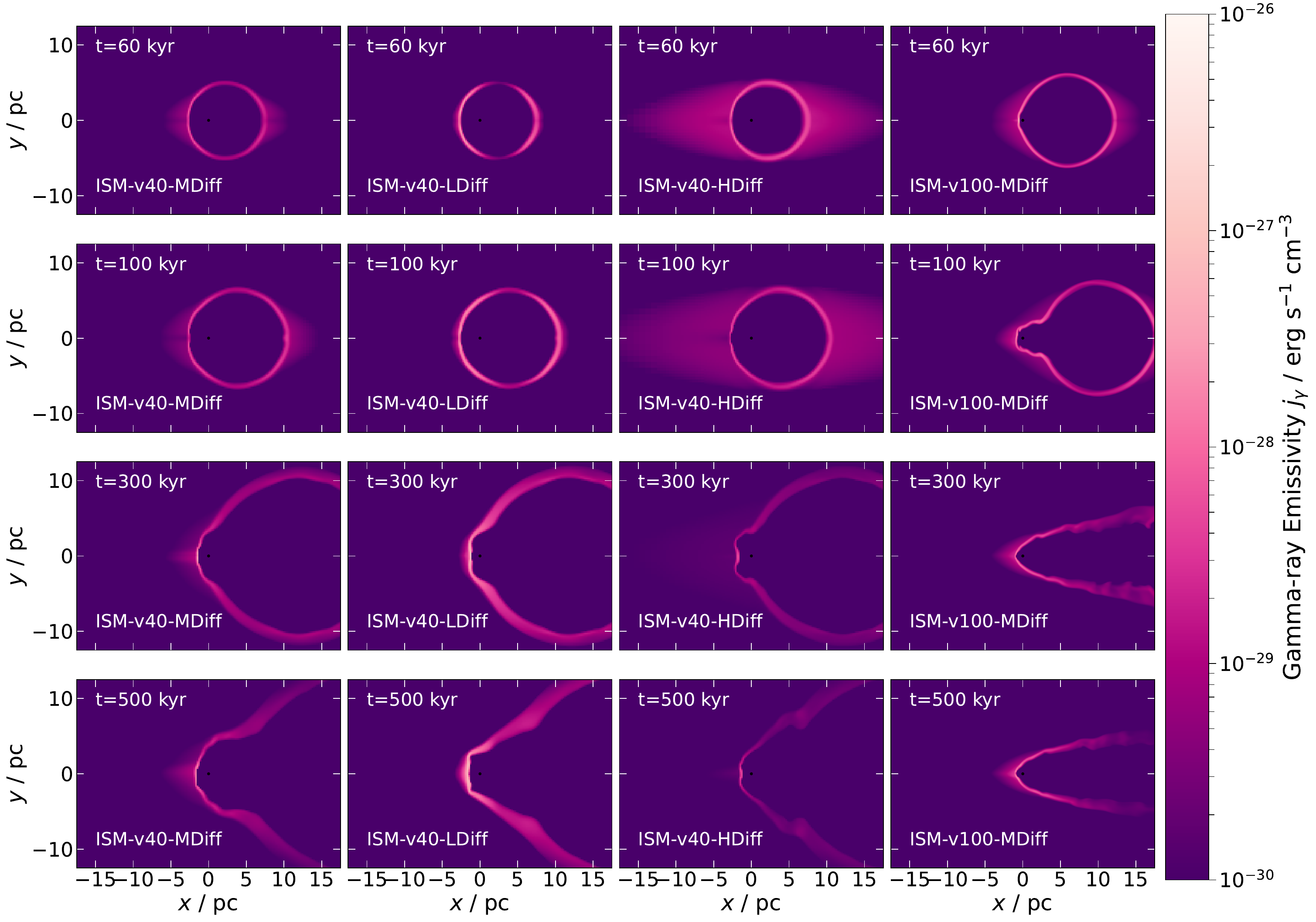}
    \caption{2D slices of the upper limits of the gamma-ray emissivity $j_\gamma$ for (from left to right) the \texttt{ISM-v40-MDiff}, \texttt{ISM-v40-LDiff}, \texttt{ISM-v40-HDiff}, and \texttt{ISM-v100-MDiff} models evaluated from this work. The slices are taken at $t = 60, 100, 300,$ and $\SI{500}{\kilo\year}$.}
    \label{fig:gamma_emission_2dslice}
\end{figure*}

\Cref{fig:gamma_emission_2dslice} show the 2D slice of the time evolution of the upper limits of the gamma-ray emissivity, $j_\gamma$, for each model considered in this work. It is evident that only the diffusion coefficient plays a major role in the gamma-ray emission from our model, as the results for \texttt{ISM-v40-MDiff} and \texttt{ISM-v100-MDiff} behave similarly, aside from morphological differences due to their difference dynamical timescales. With a low diffusion coefficient, the majority of CRs remain within the forward shock, which is also where the majority of thermal protons exist (as observed by the high density in this region). This allows for a higher interaction rate and subsequently more gamma-ray emission. With higher diffusion coefficients, this effect is reduced as there are fewer CRs at the shock front. No notable hadronic gamma-ray emission is seen near the termination shock, even with a sufficient CR population within the wind bubble (as seen in \Cref{fig:encr_timeevo_2d}), as the low magnitude of the gas density reduces the rate of proton-proton interactions near the termination shock.

\begin{figure}[!h]
    \centering
    \resizebox{\hsize}{!}{\includegraphics{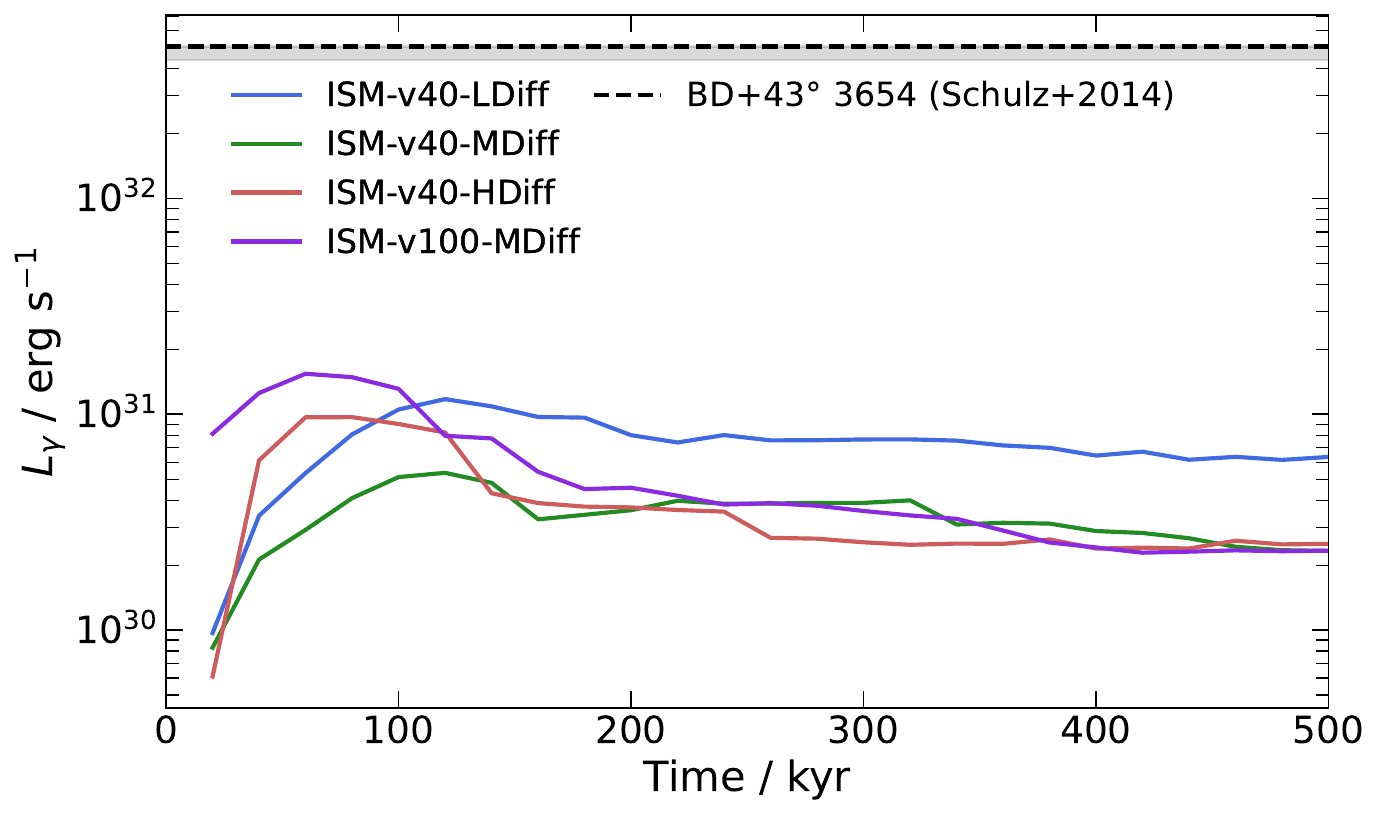}}
    \caption{Time evolution of the upper limits of the gamma-ray luminosity for each model following \Cref{eq:gamma_lum}. The dashed lines indicate upper limits of the gamma-ray luminosity as observed from \cite{SchAckBue14} averaged over the observed photon frequencies.}
    \label{fig:gamma_lum_timeseries}
\end{figure}

In \Cref{fig:gamma_lum_timeseries}, we also show the time evolution of the upper limits of the gamma-ray luminosity for the simulations considered in this work. We overlay the upper limit values of the gamma-ray luminosity of the bow shock observed from the runaway star BD+43 3654, with associated stellar mass of $\sim \SI{70}{\solarmass}$ at an age of $\sim \SI{1.5}{\mega\year}$ \citep{BenRomMar10}, averaged over the flux in each observed photon energy range. This observed gamma-ray luminosity is computed using the measured upper limits on the flux from \cite{SchAckBue14} through the conversion factor $L_\gamma = 4\pi d^2 F_\gamma$ for each energy band (assuming omnidirectional emissivity from the source) using the distance of the bow shock from Earth $d = \SI{1.72 \pm 0.03}{\kilo\parsec}$ as given in \cite{Gaia2021}. As we lack the necessary information to uncover the photon energies in which we determine the gamma-ray luminosity, we assume that the obtained luminosities are averaged over all energy bands. To match this assumption, we also average the observed upper limits on the luminosity of BD+43 3654 over all frequency bands provided in \cite{SchAckBue14}. We observe that the increase of gamma-ray production from CR protons occurs more rapidly with the \texttt{ISM-v100-MDiff} and \texttt{ISM-v40-HDiff} cases, as the smaller dynamical timescales and diffusion timescales (respectively) allow for a higher rate of CR production. Nevertheless, all models aside from the \texttt{ISM-v40-LDiff} model approach similar values, saturating at around $L_\gamma \sim \SI{3e30}{|erg\per\second}$. This indicates that, here,  thermal protons originating from the gas density is the primary driving mechanism of the production rate of gamma-rays. Only the \texttt{ISM-v40-LDiff} model retains a slightly higher luminosity value of $L_\gamma \sim \SI{6e30}{|erg\per\second}$, as the low CR diffusion strength likely allows a higher gamma ray production from CR protons. The luminosity from all models is nevertheless at least 1.5 orders of magnitude lower than the current limits from BD+43 3654. This is expected, as the stellar mass used in this work is three times smaller than those of BD+43 3654. \par 

\subsection{Radio synchrotron emission}  \label{subsec:synch_emis}

As mentioned in \Cref{sec:introduction}, radio synchrotron emission from several massive runaway stars have been observed to this date (e.g. \citealt{BenRomMar10}, \citealt{MouMacCar22}, \citealt{VanSaiMoh22}, \citealt{vandenEijnden2024}). Furthermore, the synchrotron emission from runaway stars are proposed as a possible origin for non-thermal radio filaments observed near the Galactic centre (e.g. \citealt{YusefZadeh2019}). As such, it is of most interest to estimate the synchrotron emission originating from CR electrons accelerated through the bow shocks considered in our simulation and compare our results with the observed synchrotron luminosity. \par 

The synchrotron emissivity, $j_{\mathrm{syn}}(\vec{x}, \nu) = E_\nu \diff N_e / (\diff t \diff \nu \diff ^3 x)$, at a given synchrotron frequency $\nu$ is directly related to the integral of the total (primary + secondary) CR electron population spectrum. However, we only obtain the momentum-integrated CR proton energy density, $\epsilon_\mathrm{CR}$, from our simulation. As such, we calculate the steady-state spectrum for both CR protons and electrons (primary + secondary) in momentum space, considering only the momentum-dependent losses and continuous injection of CRs. These spectra, when integrated over, allow us to formulate a relation to directly translate the CR proton energy density to the synchrotron emissivity. \par

The steady-state spectrum, within this assumption, can be solved analytically, yielding (see \Cref{app:steady_state} for more details on the calculation of the steady-state spectrum):
\begin{align}
    f_\mathrm{p}^0(t, \vec{p}_p) & = \dfrac{A_p \, p_\mathrm{max, p}^{-\alpha_\mathrm{inj}(\vec{x}) + 1}}{|b_p(\vec{p}_p, \vec{x})|} \left(\frac{|\vec{p}_p|}{p_\mathrm{max, p}}\right)^{-2} \, 
      \Gamma (3 - \alpha_\mathrm{inj}(\vec{x}), |\vec{p}_p| / p_\mathrm{max, p}), \\
    f_\mathrm{e}^0(t, \vec{p}_e) & = \dfrac{A_e \, p_\mathrm{max, e}^{-\alpha_\mathrm{inj}(\vec{x}) + 1}}{|b_e(\vec{p}_e, \vec{x})|} \left(\frac{|\vec{p}_e|}{p_\mathrm{max, e}}\right)^{-2} \, 
      \Gamma (3 - \alpha_\mathrm{inj}(\vec{x}), |\vec{p}_e| / p_\mathrm{max, e}),
    \label{eq:steady_state}
\end{align}
where $A_p, A_e$ are normalization constants with units of $\si{\per\centi\meter\cubed\per\second}$, $\Gamma(a, b)$ is the upper incomplete gamma function, $p_\mathrm{max, p}$, $p_\mathrm{max, e}$, is the maximal normalised momentum, $b_p, b_e$ represent the momentum-dependent energy loss terms for CR protons and electrons, respectively. The superscript indicates that our spectra are unnormalised since this information is encoded through the CR energy density in each cell from our simulated results. The maximal momenta for each specie is determined by equating the acceleration timescale with the loss timescale (including diffusion, advection, and losses), using parameters determined at the termination shock for all cells (see \Cref{subsec:maximal_energy} for more details). \par 

 For CR protons, we consider hadronic losses from interactions with thermal protons in the medium that dominate above the pion threshold energy $E > E_\mathrm{th} = \SI{1.22}{\giga\electronvolt}$, which competes with ionization losses due to Coulomb interactions at lower energies. For CR electrons, we consider radiative losses due to synchrotron emission and inverse Compton scattering, as well as Coulomb losses. For both CR proton and electrons, we choose an injection index of $\alpha_\mathrm{inj, p} = \alpha_\mathrm{inj, e} = 4.2$, motivated through phenomenological models for primary and secondary CR population of protons and electrons in the Milky Way \citep[e.g.][]{Moskalenko1998, Werhahn2021}. The inclusion of the time- and space-dependent injection index can be performed with a spectrally resolved algorithm. \par

As we assume that the entire CR population within our simulations is protons, we need to determine an appropriate normalisation for the electron steady-state spectrum. As such, we normalise the steady state spectrum with the observed total (primary + secondary) electron-to-proton ratio $K_{ep}^\mathrm{obs} \approx 0.01$ \citep{Cummings2016} at $\SI{10}{\giga\electronvolt}$, i.e.,
\begin{equation}
    f_e(p_e) = K_{ep}^\mathrm{obs} \frac{f_p^0(p_{p, \SI{10}{\giga\electronvolt}})}{f_e^0(p_{e, \SI{10}{\giga\electronvolt}})} \left(\frac{m_e}{m_p}\right) f_e^0(p_e),
    \label{eq:electron_spect}
\end{equation} 
where the factor $m_e / m_p$ is required due to the transformation of proton to electron momentum bins. Different CR electron population (primary or secondary) can have varying normalisation values and injection indices due to their acceleration mechanism (directly at shocks or through pion decays from CR protons, which can experience further re-acceleration at shocks). For simplicity, however, we consider both primary and secondary electrons in the description of the spectrum here, and opt for an explicit treatment of different CR electron populations in future works. With this, we can use the normalisation values determined from the CR energy density within our simulations to then normalise the energy density of CR electrons.

\subsubsection{Synchrotron emissivity}

The synchrotron emissivity $j_\mathrm{syn}(\vec{x}, \nu)$ at a given frequency $\nu$ can be described by:
\begin{equation}
    j_\mathrm{syn}(\vec{x}, \nu) = \frac{\sqrt{3} e^3 B_\perp(\vec{x})}{m_e c^2} \int_0^\infty \diff p_e 4 \pi p_e^2 \, f_e(p_e) F(\nu / \nu_c),
    \label{eq:synch_emissivity}
\end{equation}
where $B_\perp$ is the magnetic field transverse to the line-of-sight, $\nu_c(p_e, B_\perp(\vec{x}))$ is the critical frequency, and $F(\nu / \nu_c)$ is the dimensionless synchrotron kernel proportional to the integral over a modified Bessel function (\citealt{Pfrommer2022}, \citealt{Werhahn2021}). In practice, we calculate the synchrotron kernel using an analytical approximate form (see \citealt{Aharonian2010} for more details). In our model, we assume an isotropic projection so that $B_\perp^2(\vec{x}) = 2/3 |\vec{B}(\vec{x})|^2$.\par 

\begin{figure}[!h]
    \centering
    \resizebox{\hsize}{!}{\includegraphics{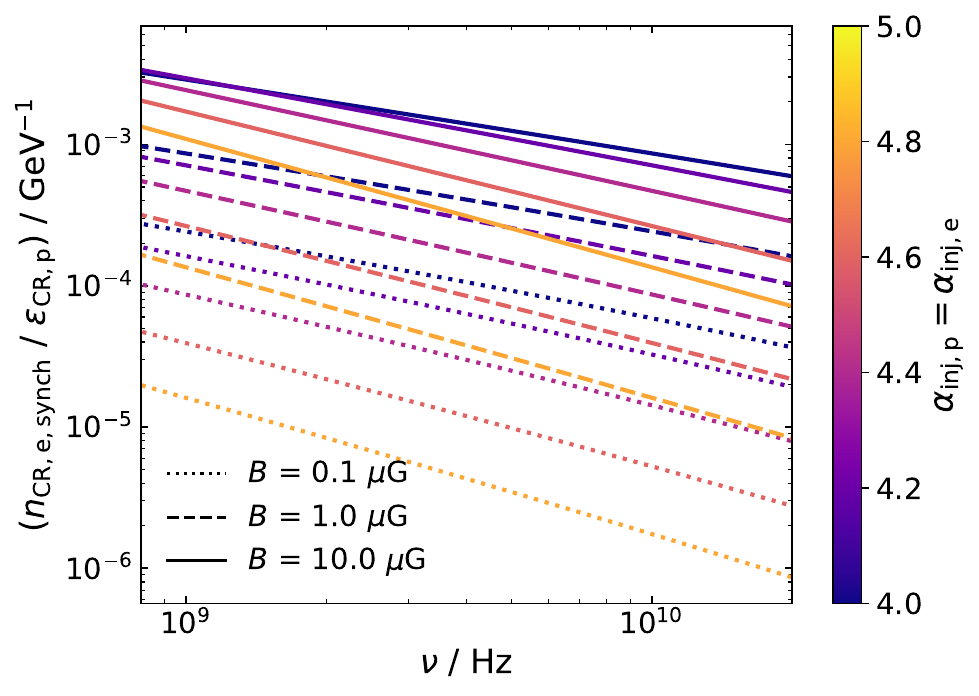}}
    \caption{Ratio of the number density of CR electrons that undergo synchrotron emission with the CR proton energy density (\Cref{eq:ensynchratio}). The magnetic field strengths are chosen based on those at the outer shock, where the majority of the emission happens. The colorbar indicates different values of $\alpha_\mathrm{inj}$, which we assume to be the same between CR protons and electrons in this work. }
    \label{fig:ncrsynch_encrprot_ratio}
\end{figure}

The synchrotron kernel can be interpreted as weights for the number of CR electrons that undergo synchrotron emission in each momentum bin. With this interpretation, we can construct a relation between the CR energy density of protons with the number of CR electrons that participate in synchrotron emission: 
\begin{equation}
    \frac{n_\mathrm{CR, e, syn}}{\epsilon_\mathrm{CR, p}} = \frac{\int_0^\infty \diff p_e 4 \pi p_e^2 \, f_e(p_e) F(\nu / \nu_c)}{\int_{0}^{\infty}\diff p_p \: 4 \pi p_p^2 T_p(p) f_p(p_p) },
    \label{eq:ensynchratio}
\end{equation}
through the proton and electron steady state spectrum defined in \Cref{eq:steady_state} and \Cref{eq:electron_spect}, respectively. Using the CR energy density from our simulation, we can use this equation to then estimate the synchrotron emissivity with \Cref{eq:synch_emissivity}. \Cref{fig:ncrsynch_encrprot_ratio} shows the dependence of this ratio on varying synchrotron frequencies and magnetic field strengths. The figure highlights that, with our prescription, weaker and higher values of the magnetic field strength and synchrotron frequency, respectively, produce fewer CR electrons that participate in synchrotron emission. The synchrotron luminosity can then be directly calculated by taking the volume integral of the synchrotron emissivity. Again, while we have implemented an exponential cutoff to suppress CR electrons at higher energies, without an energy-dependent treatment of the CR injection, we assume that all CR electron population involved can produce radio synchrotron emission at GHz frequencies, which may not hold true. As such, considering that the spectra for synchrotron electrons may cut off at lower energies, we also treat the results produced here as upper-limits of the radio synchrotron emission. \par 

\begin{figure*}[!h]
    \centering
    \includegraphics[width=17cm]{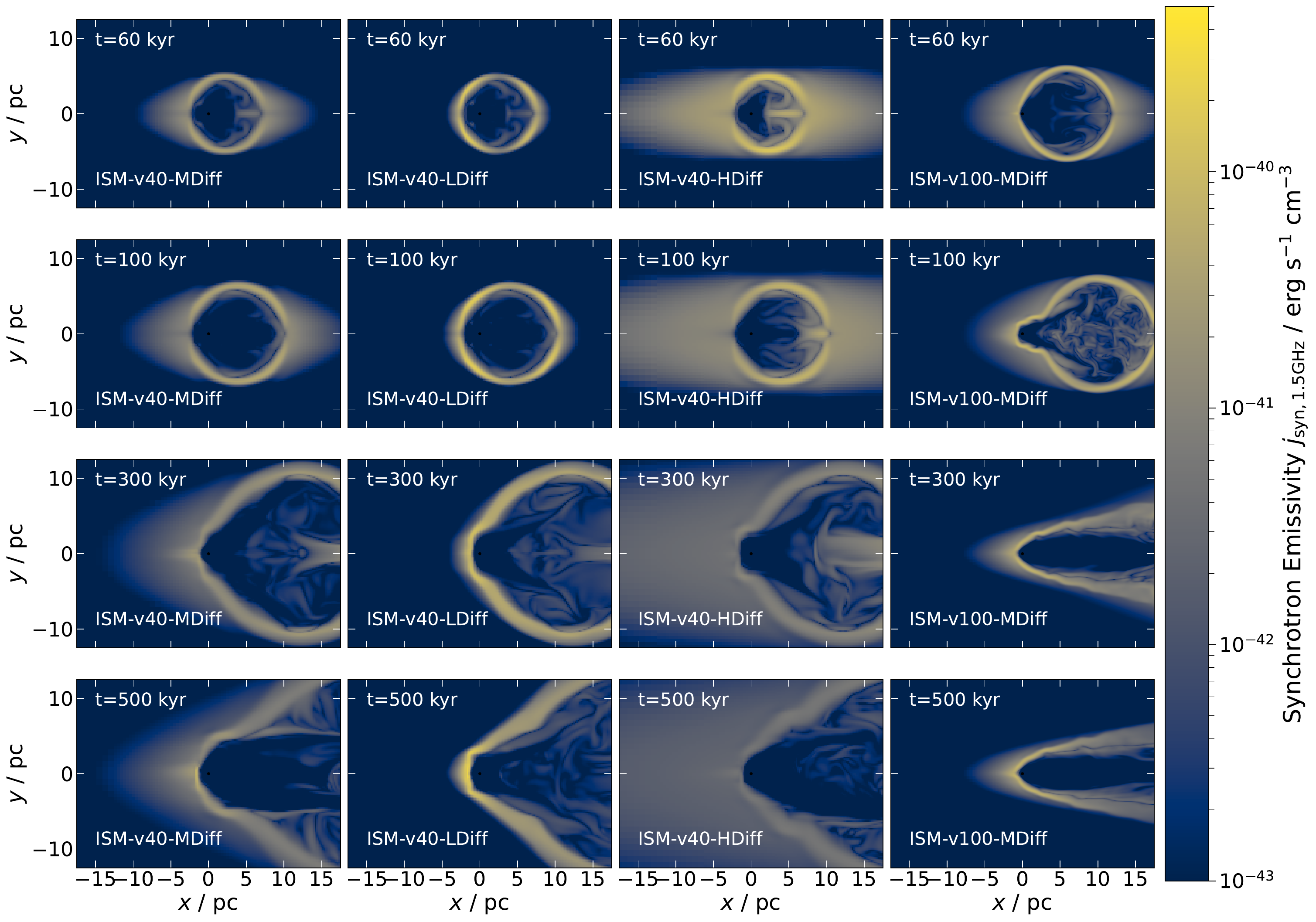}
    \caption{Same as \Cref{fig:gamma_emission_2dslice} but for the synchrotron emissivity $j_\mathrm{syn}$, evaluated at $\nu = \SI{1.5}{\giga\hertz}$, instead.}
    \label{fig:synch_emission_2dslice}
\end{figure*}

\begin{figure}[!h]
    \centering
    \resizebox{\hsize}{!}{\includegraphics{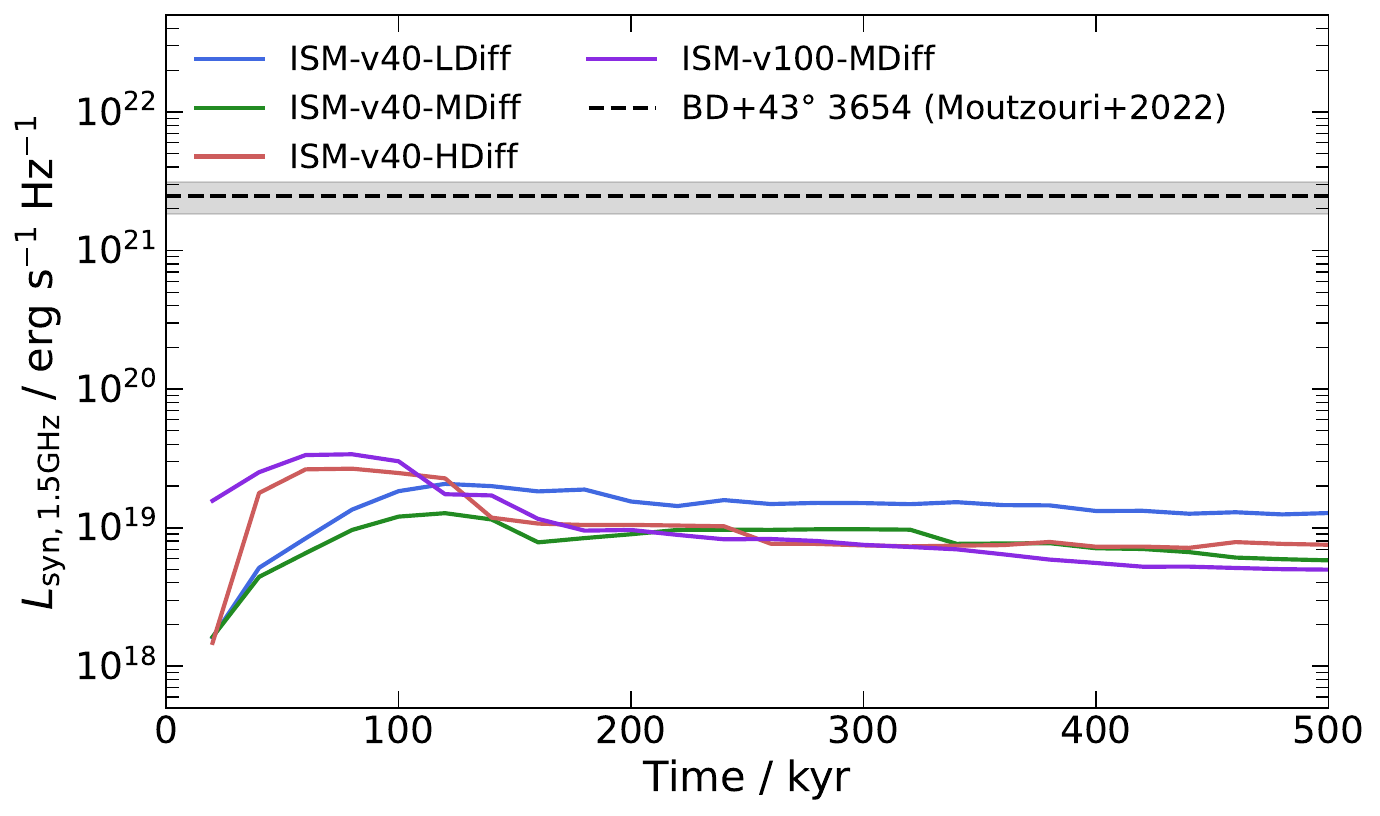}}
    \caption{Time evolution of the spectral synchrotron luminosity (in $\si{\erg\per\second\per\hertz}$) for each model by taking the volume integral of the synchrotron intensity as in \Cref{eq:synch_emissivity} at $\nu = \SI{1.5}{\giga\hertz}$.}
    \label{fig:synch_lum_timeseries}
\end{figure}

\Cref{fig:synch_emission_2dslice} show the time evolution of the upper limits of the synchrotron emissivity, $j_\mathrm{syn}$, for each model considered in this work. The emissivity is evaluated through \Cref{eq:synch_emissivity} at $\nu = \SI{1.5}{\giga\hertz}$. Similar to the gamma-ray emission, we observe that the majority of the emission occurs at the outer shock. We also observe that the CR diffusion rate again plays a major role in the overall emission, as the emission profiles between the \texttt{ISM-v40-MDiff} and \texttt{ISM-v100-MDiff} runs behave similarly. However, we notice that the morphology of the magnetic field also drives the emission profile, where the enhanced regions of synchrotron emission coincide with regions of high magnetic field strengths. As we do not include a stellar magnetic field in the simulation setup, the large magnetic field strengths with $B_\star \gtrsim \SI{10}{\gauss}$ \citep{Walder2012} can increase the overall emission from CRs accelerated at the termination shock. As such, it is to note that the synchrotron emission from the termination shock is underestimated, and can only be resolved with the inclusion of a stellar magnetic field. \par  

\Cref{fig:synch_lum_timeseries} show the time evolution of the upper limit of the spectral synchrotron luminosity at $\nu = \SI{1.5}{\giga\hertz}$ for all simulations performed in this work. We also overlay the synchrotron luminosity from BD+43 3654 at the same synchrotron frequency. The synchrotron luminosity is evaluated similarly as in \Cref{subsec:gamma_ray}, taking the synchrotron flux as reported in \citet{MouMacCar22}.  We observe a similar behaviour to \Cref{fig:gamma_lum_timeseries}, where the initial production rate of synchrotron photons is stronger for the \texttt{ISM-v100-MDiff} and \texttt{ISM-v40-HDiff} cases. However, we observe that the steady-state behaviour is largely dominated by the CR diffusion, as the luminosity strength is inversely correlated with the CR diffusion strength. Furthermore, the luminosity of \texttt{ISM-v40-MDiff} and \texttt{ISM-v100-MDiff} cases approach similar values at later times. Nevertheless, their values saturated at $L_\mathrm{syn} \sim \SI{1e19}{\erg\per\second\per\hertz}$, which is at least an order of magnitude lower than the observed luminosity from BD+43 3654. This is again expected, as the wind mechanical luminosity of the bow shock is much larger than that in our model. \par 

\begin{figure}[!h]
    \centering
    \resizebox{\hsize}{!}{\includegraphics{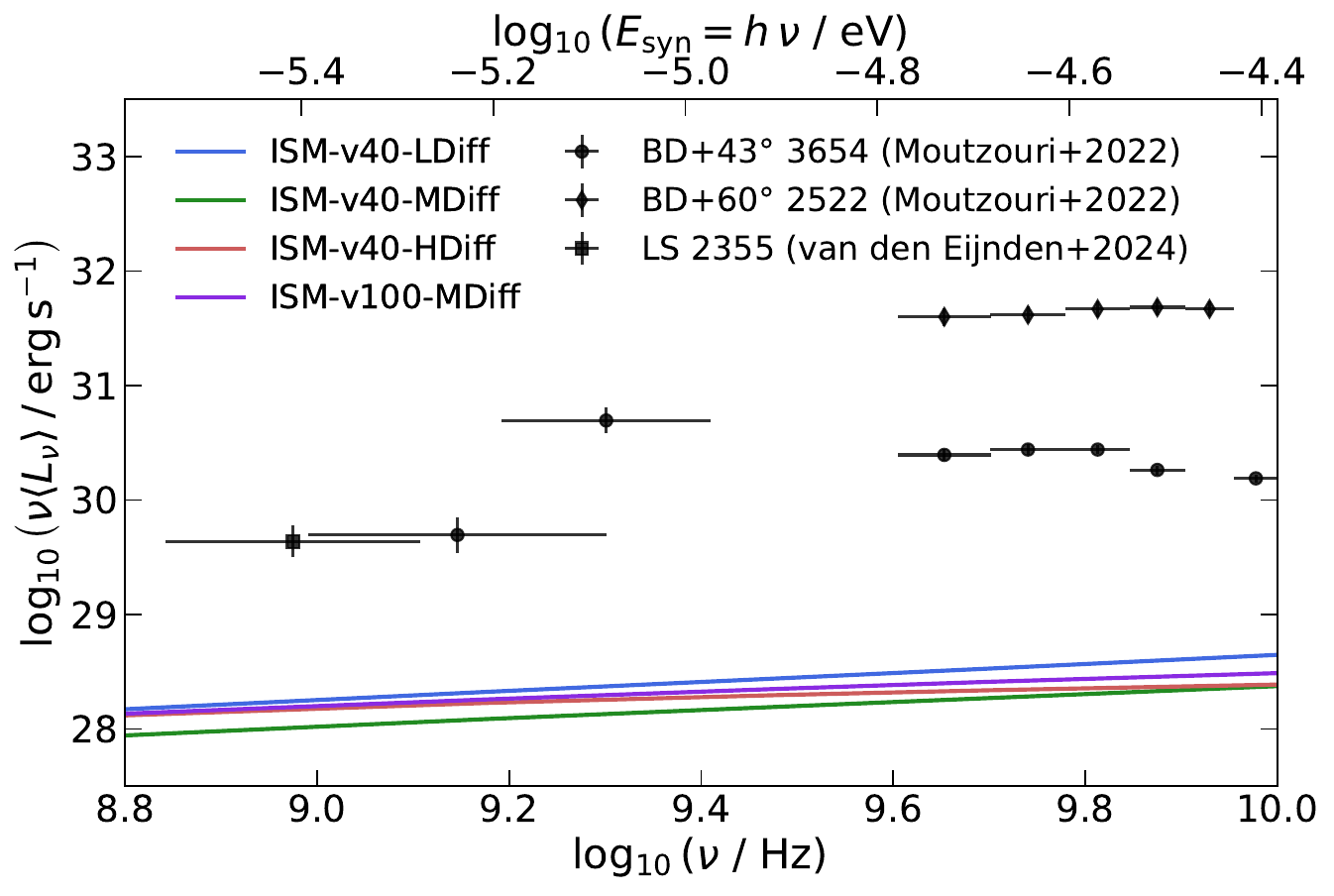}}
    \caption{Time-averaged synchrotron luminosity from each model at each synchrotron frequency. We also plot the observed synchrotron luminosity from the massive runaway stars BD+43 3654, BD+60 2552 \citep{MouMacCar22} and LS 2355 \citep{vandenEijnden2024} at each observational frequency. The top axis shows the equivalent photon energy.}
    \label{fig:synch_lum_freq}
\end{figure}

In \Cref{fig:synch_lum_freq} we directly compare our results with radio synchrotron emission from the massive runaway stars BD+43 3654, BD+60 2552 \citep{MouMacCar22} and LS 2355 \citep{vandenEijnden2024} by plotting the time-averaged synchrotron luminosity at each frequency. The synchrotron luminosities from the two runaway stars are computed in a similar fashion to those in \Cref{subsec:gamma_ray}. For BD+60 2552 and LS 2355, we use the distance from the star to Earth $\SI{3.0 \pm 0.2}{\kilo\parsec}$ \citep{Gaia2021} and $d = \SI{2.2 \pm 0.1}{\kilo\parsec}$ \citep{BailerJones2012}, respectively, to convert from the observed fluxes in \cite{MouMacCar22} and \cite{vandenEijnden2024} to the luminosities at each observed frequency. \par 

We see that the slope of the frequency dependence of the synchrotron emission $(\nu L_\nu)^a$ from our prescription is mostly positive, yielding values from 0.2 - 0.4. This is compatible with typical spectral energy distributions (SEDs) for synchrotron emission from shock acceleration, which have values of $a$ ranging from 0.2 - 0.8 (e.g. \citealp{Ranasinghe:2023}). \par 

We observe that, at all values of the observed frequency, we are consistently underestimating the emission strength from those of the observations. This is indeed expected, as the wind mechanical luminosities of BD+43 3654, BD+60 2552, and LS 2355 are approximately 250, 25, and 36 times stronger than those set from our simulations, respectively. With these strong shocks, we would observe more emission than predicted from our simulated results. As such, while it is non-trivial to scale our results to more massive stars with stronger winds, moving within a denser ISM, we would expect that stronger winds would lead to more non-thermal emission. Furthermore, the lack of stellar magnetic fields reduces the overall non-thermal emission from the termination shock, 
yielding an underestimated emission strength.

    \section{Discussion}
    \label{sec:discussion}
    \Cref{fig:gamma_lum_timeseries} and \Cref{fig:synch_lum_freq} show that we underestimate the gamma-ray upper limits and radio synchrotron emission from currently observed bow shocks. This underestimation is primarily attributed to the different kinematic luminosities as compared to the observed bow shocks, but also to the lack of stellar magnetic fields, which underestimates the synchrotron emission at the termination shock. We have made some further simplifications, in particular in the treatment of the acceleration efficiency of different CR populations, which can only be mitigated with a spectrally resolved model. Several improvements towards our simulation could make our modelling more realistic, which we list below. 

\begin{enumerate}
    \item Spectrally resolved CR model: Without a spectrally resolved CR model, we cannot determine the maximum energy of CRs, and as such even CRs accelerated at low-velocity shocks ($\mathcal{M}_1 \lesssim 3$) can contribute significantly towards particle acceleration (see \Cref{app:mach_eff}). This can lead to an over-optimistic efficiency for particle acceleration, which in turn yields optimistic gamma-ray or radio synchrotron emission. Indeed, \cite{Martinez:2023} show that a hard spectrum for $E_\mathrm{CR} \leq \SI{2}{\giga\electronvolt}$ with a maximum energy cutoff of $\gtrsim \SI{100}{\tera\electronvolt}$ is necessary to describe the non-thermal emission from observed bow shocks. Furthermore, we cannot fully utilise the cell-specific injection spectral index as we assume an integrated CR energy within each cell, which does not allow the information of the injection to be preserved after each timestep. With a spectrally resolved model, we can freely vary the slope of the CR spectrum within each cell, yielding varying CR populations that can be evaluated on-the-fly with our injection algorithm (see e.g. \cite{Girichidis2024} for a spectrally resolved CRMHD model). The power-law dependence of the injection spectrum will limit CR population with high energies, which may reduce the contributing CRs for gamma-ray / radio synchrotron emission. In particular, the particle distribution accelerated at the termination shock may be harder, while that of the outer shock may be softer (as can be seen in \Cref{fig:crshockparams_outer} and \Cref{fig:crshockparams_term}), which may enhance emission from the termination shock and reduce that of the outer shock. Through this approach, we can also directly incorporate momentum dependence in CR transport parameters such as from diffusion or Coulomb losses, which is currently not implemented within the grey approach.
    \item Inclusion of a stellar magnetic field: In our current model, we assume that no magnetic fields are generated from the star. However, the stellar rotation can induce a toroidal component of the magnetic field, which decays $\sim 1 / R$ at large distances. As such, the magnetic field at the termination shock can be produced from either the adiabatic compression of the stellar magnetic field or from in-situ amplification through the Bell instability \citep[e.g.][]{DelBosMul18, delValle2018}. This, along with efficient CR acceleration at the termination shock, will increase the synchrotron emission observed at the termination shock. While including a stellar magnetic field is beyond the scope of this work, it will be of interest to observe the impact of the non-thermal emission with its inclusion.
    \item Improved CR acceleration model: Our current model for the dependence of the Mach number on the CR acceleration efficiency is based on a fitting function produced from \citet{Kang2007}. Utilising this function, coupled with a maximal CR acceleration efficiency motivated from current studies, is a good first estimate to model the CR acceleration. However, to fully incorporate an accurate modelling of the CR acceleration efficiency, it would be beneficial to utilise tabulated values from hybrid PIC simulations that can additionally include and consider the dependence on thermal-to-CR pressure, which has been shown to further impact the CR acceleration efficiency \citep{Kang2013}. The indistinguishability between low- and high-velocity shocks can also influence the overestimation of particle acceleration at shocks (see \Cref{app:mach_eff}). The amplification of magnetic fields due to CR streaming is also crucial for efficient CR acceleration with the DSA framework. An implementation of shock acceleration with CR pressure dependence and the inclusion of CR streaming \citep{Dubois2019} within grid-based architectures have already been developed and, as such, can be easily translated into our framework.
    \item Inclusion of primary electrons accelerated at shocks: Currently, the CR electron population is determined indirectly through the integrated CR electron steady-state spectrum, which includes both primary and secondary CR electrons. However, including an additional contribution of CR electrons into the overall CRMHD equations will not only allow primary CR electrons to be accelerated through shocks with our current implementation, but also more easily (and directly) retrieve the synchrotron luminosity without any assumption of the spectrum. Incorporating this with a spectrally resolved CRMHD model (as stated in point 1) will further improve on the modelling of the overall electron population. 
    \item Distinguishing between primary and secondary electrons: Here, the total CR electron population is assumed to contribute similarly to the overall synchrotron luminosity. However, previous works have already shown that the spectral indices and normalisation of primary and secondary electrons can differ \citep[e.g. ][]{Werhahn2021}. As such, treating each population separately is crucial to appropriately model the physics that can alter the resulting synchrotron emission. With the inclusion of primary electrons (point 4) and a spectrally resolved model (point 1), we can directly evaluate the expected secondary electron population e.g. through pion decay cross sections.
    \item Inclusion / Distinction of radiative shocks: In this current implementation, we only treat shock acceleration through non-radiative shocks. As such, we assume that the total dissipated energy is entirely attributed through such non-radiative shocks. However, with multi-source systems (such as in stellar clusters or near the Galactic centre with multiple supernovae) or for realistic systems that include other feedback processes, the energetics through radiative cooling and heating can have a non-negligible impact on the overall shock physics, which can modify the overall injected CR energy per cell. By including radiative shocks as an additional contribution to the overall dissipated energy, we can apply our implementation towards more realistic astrophysical scenarios.
    
\end{enumerate}

    \section{Conclusion}
    \label{sec:conclusions}
    In this work, we perform 3D ideal CR-MHD simulations in the advection-diffusion limit to simulate the early evolution of wind-driven bow shocks generated from massive runaway stars. We implemented a numerical scheme in the Eulerian grid-based MHD code \texttt{FLASH} following \cite{Pfrommer2017} to incorporate dynamic injection of cosmic rays through diffusive shock acceleration. The injection region is determined through an on-the-fly shock detection algorithm using gradient-based criteria and the CR energy injected in each cell is calculated utilising a semi-analytical CR acceleration efficiency. The implementation was verified through a series of standard numerical tests, which show excellent agreement with analytical predictions. \par 

We performed simulations assuming an ambient ISM displaced by the stellar wind of a main-sequence star of $\SI{20}{\solarmass}$ for five different configurations, namely three cases with $v_\star = \SI{40}{\kilo\meter\per\second}$ at different strengths of diffusion coefficients, one case with a higher stellar velocity of $v_\star = \SI{100}{\kilo\meter\per\second}$, and a reference model without any CR injection. We predict that, through dynamic injection of cosmic rays, the morphology of the bow shock is modified due to the impact of the additional CR pressure. With a higher CR diffusion rate, more CRs are diffused away from the bubble, and the dynamics of the bow shock approaches that of the MHD case. However, we observe that the majority of the strength of the magnetohydrodynamic parameters at the shocked regions are not significantly affected by the addition of CRs. We also showed that the CR energy and effective adiabatic index vary along the polar angle of the bow shock, an effect which decreases with stronger diffusion. In all cases, the termination shock proves to be a constant region for efficient particle acceleration, however that may be over-optimistic for slow shocks due to the nature of our implementation. With a higher stellar velocity, the morphological differences are primarily driven by the stellar kinematics, and we do not observe significant deviations in the CR energy and effective adiabatic index. However, we observe that the forward region of the outer shock for the \texttt{v100} case at later times can also efficiently accelerate CRs.  \par 

To compare our simulated results to observations, we calculate the gamma-ray and radio synchrotron emission through a simplified approximation. For the synchrotron emission, we first calculate the electron steady-state spectrum through a simple model with only injection and energy-dependent losses while incorporating the observed electron-to-proton ratio for normalisation. As we cannot distinguish between CR populations with high and low energies, we assume that all CR electrons and protons have high enough energies to contribute to pion-decays ($\gtrsim \SI{100}{\mega\electronvolt}$) and synchrotron emission ($\sim \si{\giga\electronvolt}$), which may be over-optimistic for acceleration at the forward shock. With our simplified spectral model assuming a uniform CR proton and electron injection index, we show that the CR diffusion has a significant impact on the time evolution of both $\gamma$-ray and radio synchrotron emission. While we assume efficient particle acceleration, we underestimate the resulting emission by at least one order of magnitude. This is due to the lower wind luminosity from our models as well as the lack of a stellar magnetic field, which can enhance the resulting non-thermal emission from the termination shock. The results obtained through this work can be further improved by including a stellar magnetic field, as well as utilising a spectrally resolved CRMHD framework with a more sophisticated CR acceleration model that can not only appropriately treat the energy distribution of CR protons after injection, but also additionally include primary CR electrons as an additional species that can dynamically evolve throughout the simulation. \par 

    \begin{acknowledgements}
        Funded by the Deutsche Forschungsgemeinschaft (DFG, German Research Foundation) – Project-ID 500700252 – SFB 1601. 
        The authors gratefully acknowledge the Gauss Centre for Supercomputing e.V. (www.gauss-centre.eu) for funding this project by providing computing time through the John von Neumann Institute for Computing (NIC) on the GCS Supercomputer JUWELS at J\"ulich Supercomputing Centre (JSC).
        JM is grateful to SW for providing use of a guest office at the University of Cologne during the preparation of this manuscript.
        SW and PCN gratefully acknowledge funding via the Collaborative Research Center 1601 (SFB 1601, subprojects B5 and B6) funded by the German Science Foundation (DFG). Visualisation of the data has been performed through the software package \texttt{flash-amr-tools}\footnote{\texttt{\url{https://pypi.org/project/flash-amr-tools/}}}. The following software were used in the analysis: \texttt{NumPy} \citep{harris2020array}, \texttt{SciPy} \citep{2020SciPy-NMeth}, \texttt{matplotlib} \citep{Hunter:2007}, \texttt{h5py}\footnote{\texttt{\url{https://www.h5py.org}}}, \texttt{astropy} \citep{Astropy2022}, \texttt{iPython} \citep{PER-GRA:2007}.  
    \end{acknowledgements}

    \bibliographystyle{aa} 

    \begin{appendix} 
    \section{CR Acceleration Efficiency as a function of the Pre-Shock Mach Number}
\label{app:mach_eq}

The dependence of the CR acceleration efficiency with respect to the pre-shock Mach number, $\xi(\mathcal{M}_1)$, is determined through a fitting function calculated from \citet{Kang2007}. Here, different functional forms are used with and without pre-existing CRs, which is determined if the CR energy density in each cell is greater than zero. Furthermore, as the acceleration efficiency in this functional form is overestimated by around 0.4, we renormalise the function such that the maximal acceleration efficiency $\xi_\mathrm{max} = 0.2$, following typical maximal values from recent hybrid PIC simulations \citep{Caprioli2014,Kang2013}. For completeness, we present the full functional form below. \par

\begin{equation}
    \xi(\mathcal{M}_1) = 
    \begin{cases}
        \quad \dfrac{\xi_\mathrm{max}}{\xi(\mathcal{M}_{1, \mathrm{max}})}\xi_\mathrm{CR}(\mathcal{M}_1) \quad & : \quad \epsilon_\mathrm{CR} > 0 \\
        \quad \dfrac{\xi_\mathrm{max}}{\xi(\mathcal{M}_{1, \mathrm{max}})}\xi_\mathrm{no CR}(\mathcal{M}_1) \quad & : \quad  \mathrm{else},
    \end{cases}
    \label{eq:craccel_fit}
\end{equation}
where:
\begin{align}
    &\xi_\mathrm{no CR}(\mathcal{M}_1) = 
    \begin{cases}
        \quad A_0 (\mathcal{M}_1^2 - 1) \quad &: \quad \mathcal{M}_1 \leq 2\\
        \quad \sum\limits_{k=0}^{4} a_k \dfrac{(\mathcal{M}_1 - 1)^k}{\mathcal{M}_1^4} \quad &: \quad \mathcal{M}_1 > 2 
    \end{cases}\quad, \\
    &A_0 = \num{1.96e-3}, \, a_0 = 5.46, \, a_1 = -9.78, \\
    &a_2 = 4.17, \, a_3 = -0.334, \, a_4 = 0.570,
    \label{eq:craccel_fit_noCR}
\end{align}
and:
\begin{align}
    &\xi_\mathrm{CR}(\mathcal{M}_1) = 
    \begin{cases}
        \quad B_0 \delta_\mathrm{th}(\mathcal{M}_1) \quad &: \quad \mathcal{M}_1 \leq 1.5\\
        \quad \sum\limits_{k=0}^{4} b_k \dfrac{(\mathcal{M}_1 - 1)^k}{\mathcal{M}_1^4} \quad &: \quad \mathcal{M}_1 > 1.5
    \end{cases}\quad, \\
    &B_0 = 1.025, \, b_0 = 0.240, \, b_1 = -1.56, \\
    &b_2 = 2.80, \, b_3 = 0.512, \, b_4 = 0.557.
    \label{eq:craccel_fit_CR}
\end{align}

Here, $\delta_\mathrm{th}$ is the gas thermalisation efficiency, which can be derived analytically from the Rankine-Hugoniot jump conditions:
\begin{equation}
    \delta_\mathrm{th}(\mathcal{M}_1) = \frac{2 \left(y_s(\mathcal{M}_1) - x_s(\mathcal{M}_1)^{\gamma_\mathrm{th}}\right)}{\gamma_\mathrm{th}(\gamma_\mathrm{th} - 1)\mathcal{M}_1^2 x_s(\mathcal{M}_1) } ,
    \label{eq:gas_therm}
\end{equation}
where the density and pressure jump ($x_s$ and $y_s$ respectively) can also be expressed entirely in terms of the pre-shock Mach number \citep{Ryu2003}.

\section{Verification of Mach number and Dissipation Energy Computation} \label{app:sod_mach}
To determine the accuracy of the evaluation of the pre-shock Mach number $\mathcal{M}_1$ and the dissipated energy flux $f_\mathrm{diss}$, we perform a series of standard Sod shock tube tests used to benchmark the performance of the solver \citep{Sod1978}. \par 

We follow a similar test performed in \cite{Schaal2015} where we consider a 2-D simulation domain $(x,y) \in [0, 100] \times [0, 50]$ with a thermal adiabatic index $\gamma_\mathrm{th} = 5/3$ for all tests. We choose the same initial parameters as in the Sod test in \Cref{subsec:sodtest}, however we modify the pressure on the left side $P_1$ of the discontinuity such that we yield a specific Mach number for each of our simulations (see \Cref{tab:sod_pleftvals}). We do not consider any CRs or magnetic fields within this test. We perform our tests until the shock reaches $x = 75$ to ensure that the shock has fully evolved. We show the time at which the measurements are performed on the right-most column of \Cref{tab:sod_pleftvals}.

\begin{table}[!h]
    \caption{Pressure on the left side of the discontinuity $P_1$ used for the initial conditions of the Sod shock tube test, chosen such that we obtain a specific pre-shock Mach number $\mathcal{M}_{1, \mathrm{thr}}$. The time in which the shock reaches $x = 75$ where we perform our measurements is also shown \citep{Schaal2015}.}
    \label{tab:sod_pleftvals}
    \centering
    \begin{tabular}{ccc}
    \hline \hline

        $P_1$ & $\mathcal{M}_{1, \mathrm{thr}}$ & $t_{x = 75}$ \\ \hline
        0.81445 & 1.5 & 14.43 \\
        1.9083  & 2.0 & 10.83 \\
        15.358 & 5.0 & 4.330 \\
        63.499 & 10.0 & 2.165 \\
        400.52 & 25.0 & 0.866 \\
        1604.15 & 50.0 & 0.433 \\
        6418.68 & 100.0 & 0.217 \\
        40120.4 & 250.0 & 0.0866 \\
        160483.9 & 500.0 & 0.0433 \\
        641937.7 & 1000.0 & 0.0217 \\ \hline

    \end{tabular}
    
\end{table}

\begin{figure}[!ht]
    \resizebox{\hsize}{!}{\includegraphics{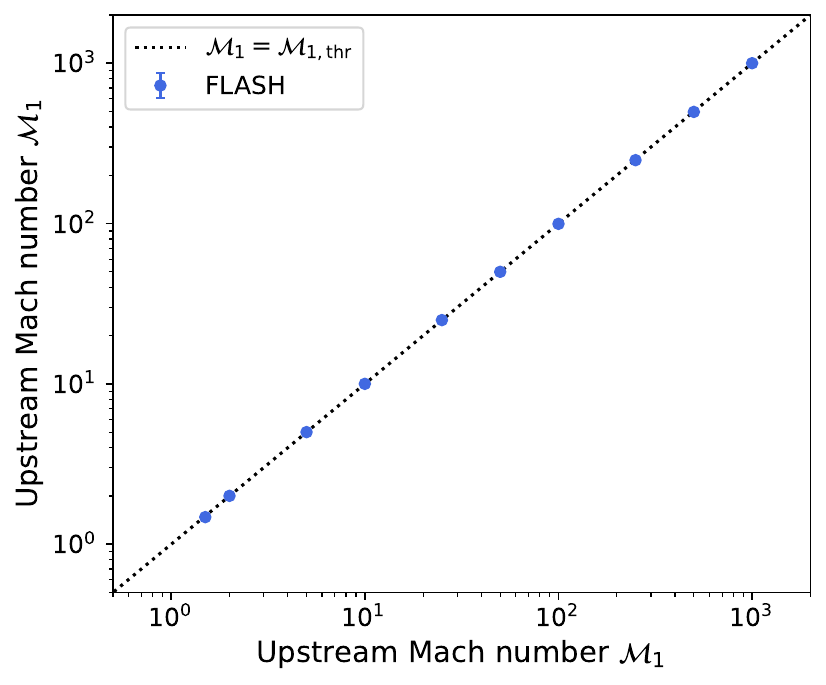}}
    \caption{The pre-shock Mach number $\mathcal{M}_1$ obtained from a series of Sod shock tube tests measured at $x = 75$. Any deviations from the linear plot are indications of deviations from the analytical values provided in \Cref{tab:sod_pleftvals}. The statistical error bars are sufficiently small such that they are not visible in this figure. }
    \label{fig:sod_mach_num}
\end{figure}

\begin{figure}[!h]
    \resizebox{\hsize}{!}{\includegraphics{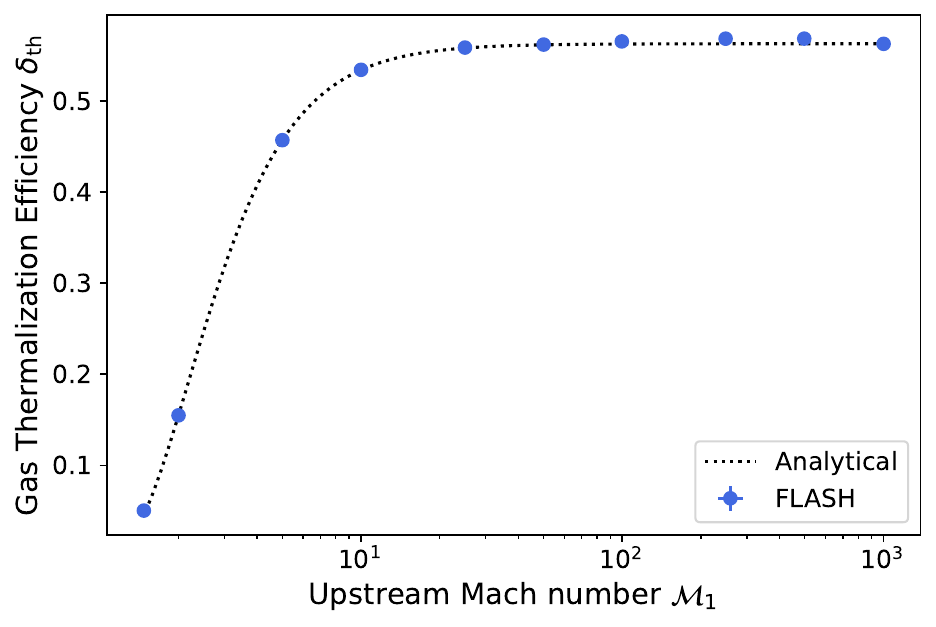}}
    \caption{Gas thermalisation efficiency $\delta_\mathrm{th}$ at different pre-shock Mach numbers $\mathcal{M}_1$ obtained from a series of Sod shock tube tests measured at times shown in \Cref{tab:sod_pleftvals}. The analytical solution in \Cref{eq:gas_therm} is also shown. The statistical error bars are sufficiently small such that they are not visible in this figure.}
    \label{fig:sod_gas_therm}
\end{figure}

In \Cref{fig:sod_mach_num}, we compare the pre-shock Mach number averaged over the $y$-direction from our simulation with the analytical values provided in \Cref{tab:sod_pleftvals}. Overall they show excellent agreement between each other.

To analyse the accuracy of the calculation of the dissipated energy, we determine the gas thermalisation efficiency within our simulation, which quantifies the amount of shock-dissipated energy that is converted to thermal energy. This is expressed by taking the ratio of the dissipated thermal flux $f_\mathrm{th}$ with the kinetic flux $f_\mathrm{kin} = 0.5 \rho_1 (\mathcal{M}_1 c_1)^3$. We compare the gas thermalisation efficiency obtained from our simulations against the analytical results as described in \Cref{eq:gas_therm} in \Cref{fig:sod_gas_therm}, where we use the Mach number obtained in \Cref{fig:sod_mach_num}. From this figure, we observe an excellent agreement with the analytical solution apart from minor deviations at $\mathcal{M}_1 = 250$. \par

\section{Contribution of Low-velocity Shocks on Particle Acceleration at the Forward Shock} \label{app:mach_eff}

\begin{figure}[!hb]
    \centering
    \resizebox{\hsize}{!}{\includegraphics{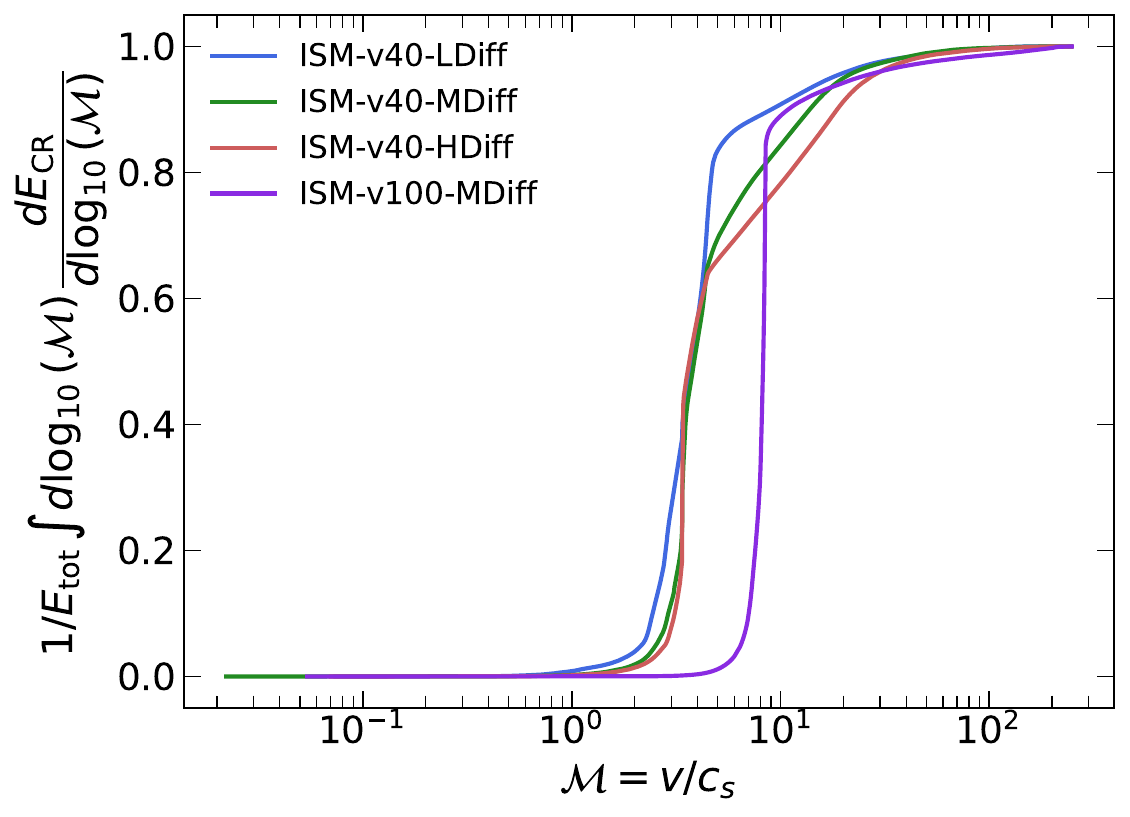}}
    \caption{Normalised cumulative distribution function of the CR energy, binned over the logarithm of the Mach number $\mathcal{M} = v / c_s$ of the magnitude of the velocity in each cell. The distribution is shown for all models used in this work at $t = \SI{500}{\kilo\year}$.}
    \label{fig:encr_mach_cumsum}
\end{figure}

Through our DSA prescription, without a spectrally resolved CRMHD model, we cannot accurately predict the energy distribution of CRs that are accelerated through our implementation. As such, CRs that are accelerated at low-velocity shocks can contribute equally to the non-thermal emission as compared to those accelerated at high-velocity shocks. In \Cref{fig:encr_mach_cumsum}, we show that more than half of CRs that are accelerated yield from low-velocity shocks of $\mathcal{M} \lesssim 10$. At these regimes, the shock physics that can accelerate particles are not well known as compared to regimes with stronger shocks. To accurately treat shocks at lower velocities, a proper prescription that not only considers the maximum CR energy available for acceleration (which can be included through a spectrally resolved model), but also one that parametrises the dependence on the shock velocity is required. 

\section{Calculation of the Steady-State Spectrum}  \label{app:steady_state}

To calculate the spectrum for both CR proton and electrons, we solve the steady-state solution of the Fokker-Planck equation resulting only from momentum-dependent cooling and continuous injection. We define the CR spectrum as the distribution of particles in 6-D phase space (3 momentum and 3 spatial dimensions) so that $f := \diff^6 N / (\diff^3 \vec{p} \diff^3 \vec{x})$. \par 
As the spatial component of the Fokker-Planck equation is solved directly in the simulations through the (integrated) CR transport equation, we necessarily only need to solve for the momentum component. We take a power-law to describe the CR injection spectrum at each spatial position $\vec{x}$, with an additional exponential cutoff depending on the maximal momentum $p_\mathrm{max}$:
\begin{equation}
    j(t, \vec{p}) = A |\vec{p}|^{-\alpha_\mathrm{inj}(\vec{x})} \, \exp(-p / p_\mathrm{max}),
\end{equation}
where $A$ is a normalization constant with units of $\si{\per\centi\meter\cubed\per\second}$, and $\alpha_\mathrm{inj}(\vec{x})$ is the spatial-dependent CR injection index, obtained through our DSA prescription. The resulting steady-state solution of the Fokker-Planck equation can be analytically solved for $\alpha_\mathrm{inj} > 3$ (without the inclusion of momentum-dependent diffusion):
\begin{equation}
    f_s^0(t, \vec{x}, \vec{p}_s) =
     \dfrac{A \, p_\mathrm{max, s}^{-\alpha_\mathrm{inj}(\vec{x}) + 1}}{|b_s(\vec{p}_s, \vec{x})|} \left(\frac{|\vec{p}_s|}{p_\mathrm{max}}\right)^{-\alpha_\mathrm{inj}(\vec{x})} \, 
      \Gamma (3 - \alpha_\mathrm{inj}(\vec{x}), |\vec{p}_s| / p_\mathrm{max, s}),
    \label{eq:steady_state}
\end{equation}
where $\Gamma(a, b)$ is the upper incomplete gamma function. The superscript indicates that our spectra are unnormalized since this information is encoded through the CR energy density in each cell from our simulated results. We compute the resulting spectrum for cosmic ray protons and electrons separately (labelled by $s \in \{p, e\}$), where separate momenta and injection indices are used for each species. \par 

We note that as the spectral information is not communicated between each timestep, the injection index obtained from our algorithm is only valid within the shock region determined at that timestep. As such, in this work, we assume a uniform injection index throughout the domain for both cosmic ray protons and electrons. Nevertheless, the time- and space-dependent injection index can be trivially included within this formalism with a spectrally resolved algorithm. \par

\begin{figure}
    \centering
    \resizebox{\hsize}{!}{\includegraphics{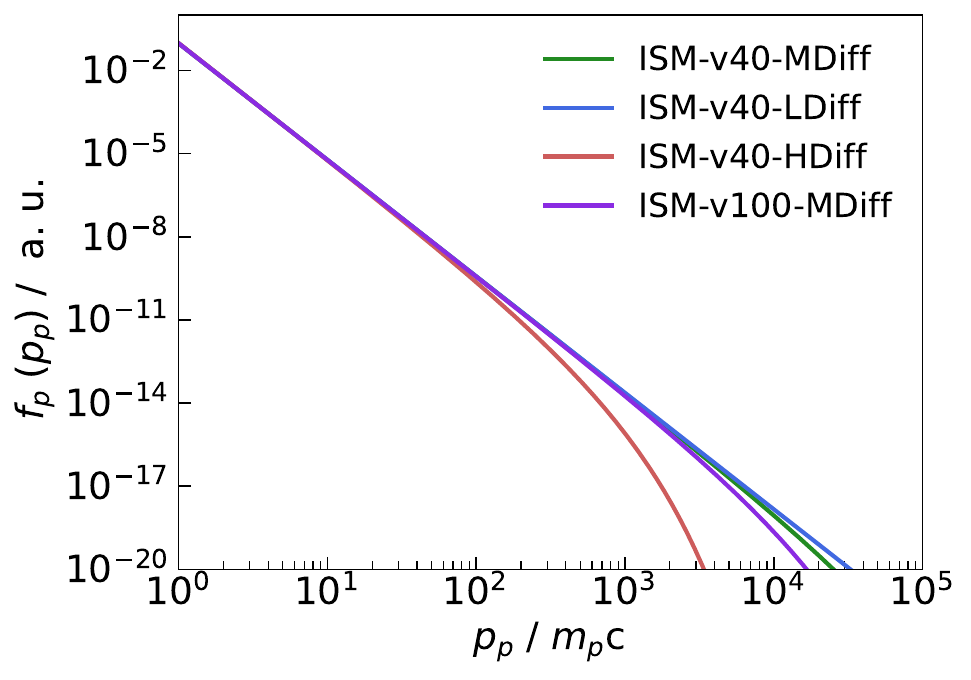}}
    \caption{Steady-state spectra for CR protons for all models used in this work. Here, we take an injection spectral index $\alpha_\mathrm{inj} = 4.2$, a gas density $\rho = \SI{1e-25}{\gram\per\centi\meter\cubed}$, and a fluid velocity of $v = \SI{100}{\kilo\meter\per\second}$, as nominal values near the shock. The behaviour of the spectra is approximately a power-law, up to a certain momentum value, where the exponential suppression, primarily driven by the CR diffusion, dictates the behaviour of the spectra at higher momenta. }
    \label{fig:steady_state_proton}
\end{figure}

\begin{figure}
    \centering
    \resizebox{\hsize}{!}{\includegraphics{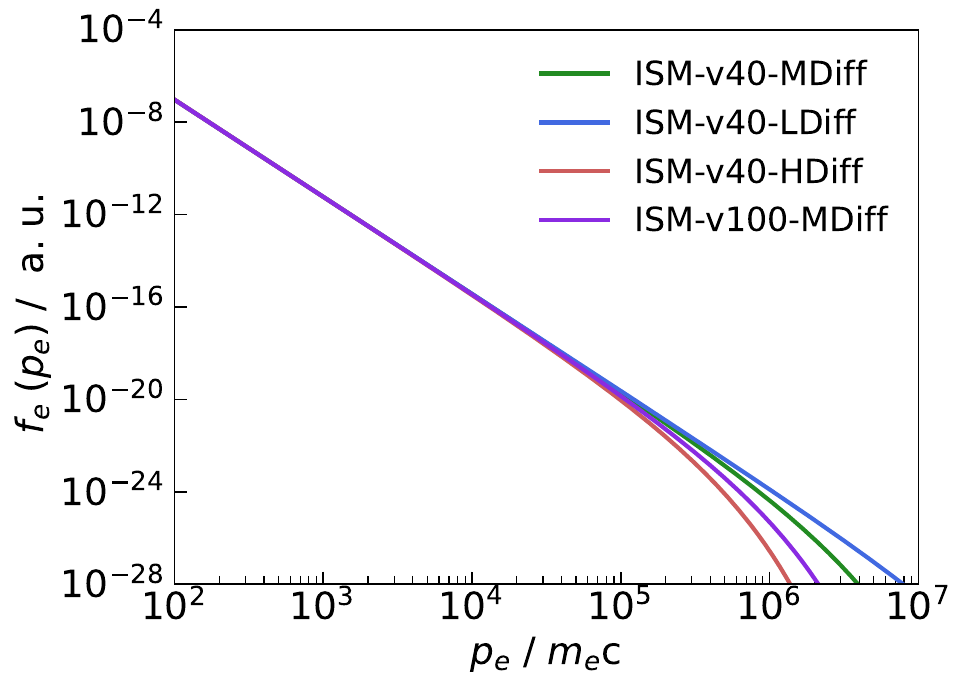}}
    \caption{Same as \Cref{fig:steady_state_proton} but for CR electrons instead. As the synchrotron losses are driven by the magnetic field strength, we take a nominal value of $B = \SI{1}{\micro\gauss}$. The features are similar as with CR protons, except that, due to the choice of the minimum grid momenta, CR electrons with $\gtrsim \SI{50}{\mega\electronvolt}$ can also contribute to the overall spectra.}
    \label{fig:steady_state_electron}
\end{figure}

The calculation of the steady-state spectrum is performed by choosing a grid of proton and electron momenta and determining the resulting spectra as shown in \Cref{eq:steady_state}. For protons, we choose a minimum grid momenta of $p_\mathrm{p, min} = \SI{1}{\giga\electronvolt} / c$, as the CR spectrum below this value no longer behaves as a power-law, thus breaking the power-law assumption of the injection spectrum. The minimum momenta of electrons is set to be $p_\mathrm{e, min} = \SI{0.051}{\giga\electronvolt} / c$ as CR electrons can yield momenta < GeV energies while retaining its power-law behaviour \citep[see e.g.][]{Cummings2016}. The maximum grid momentum is set to $\SI{9.4e4}{\giga\electronvolt} / c$ and $\SI{5.1e3}{\giga\electronvolt} / c$ for CR protons and electrons respectively, sufficiently large such that the high-energy CRs are suppressed through the exponential cutoff. In practice, we use the normalised momenta ($P_s = p_s / (m_s c)$) to allow the numerical integration to be stable. \Cref{fig:steady_state_proton} and \Cref{fig:steady_state_electron} show the resulting steady-state spectra for both CR protons and electrons, respectively, for all models used in this work. While we choose nominal values for the gas density, fluid velocity, magnetic field strength, and injection spectral index for the figures shown, in practice, the spectra is calculated for each cell position per timestep. We observe that the exponential cutoff introduces a suppression for CRs at high momenta, which is primarily driven by the CR diffusion. \par 

In the subsequent subsections, we elaborate on the calculation of the energy loss processes (for both CR protons and electrons), as well as the calculation for the maximal momenta used in the exponential cutoff.

\subsection{Energy Losses for Protons} \label{subsec:energy_loss_prot}

Protons experience hadronic losses above the pion energy threshold, i.e. above energies where pions can be produced from interactions with thermal protons. This corresponds to CR energies $E \geq E_\mathrm{th} = \SI{1.22}{\giga\electronvolt}$ or equivalently for CR proton momenta above $p_\mathrm{p, th} = \SI{0.78}{\giga\electronvolt} / c$. The resulting hadronic loss component can be described as follows \citep{Girichidis2020}:
\begin{equation}
    b_\mathrm{hadr}(p, \vec{x}) = c \sigma_{pp} K_p n_N(\vec{x}),
    \label{eq:hadr_loss}
\end{equation}
where $\sigma_{pp} \approx \SI{44}{\barn}$ is the total pion cross section \citep{Pfrommer2004} and $K_p \approx 1/2$ is the inelasticity of the interaction. The nucleon density $n_N = \rho(\vec{x}) / (\mu m_p)$ is described through the gas density $\rho$ in each cell and the mean molecular weight $\mu = 1.4$ that takes into account the population of hydrogen and helium within the medium. \par 

At energies lower than $\sim$ GeV, protons can also lose energy due to Coulomb interactions with electrons within the medium. We follow the prescription from \citet{Gould1972a} to describe the Coulomb losses for cosmic ray protons: 
\begin{equation}
    b_\mathrm{Coul, p}(p, \vec{x}) = \dfrac{3 \sigma_T n_e(\vec{x}) m_e c^3}{2 \beta} \left[\ln\left(\frac{2 \gamma m_e c^2}{\hbar \omega_\mathrm{pl}} \right) - \frac{1}{2}\beta^2\right],
    \label{eq:coul_loss_proton}
\end{equation}
where $n_e = 0.88 \rho(\vec{x}) / (\mu m_p)$ is the electron number density, $\omega_\mathrm{pl} = \sqrt{4 \pi n_e e^2 / m_e}$ is the plasma frequency, and $\sigma_T = \SI{0.665}{\barn}$ is the Thomson cross section \citep{Eite2018}. The Lorentz factor $\gamma = \sqrt{1 + p^2}$ and beta factor $\beta = p / \gamma$ are defined in terms of the normalised momentum used in this work.

\subsection{Energy Losses for Electrons} \label{subsec:energy_loss_ele}

High-energy electrons primary undergo synchrotron and inverse Compton losses. The synchrotron losses can be described by the following prescription (see e.g. \cite{Blumenthal1970}): 
\begin{equation}
    b_\mathrm{syn}(p, \vec{x}) = \frac{4}{3} \sigma_T c \left(\frac{p_e}{m_e c}\right)^2 \epsilon_B(\vec{x}),
    \label{eq:synch_losses}
\end{equation} 
where $\epsilon_B = |\vec{B}(\vec{x})|^2 / 8\pi$ is the magnetic energy density in each cell. \par

Both the interstellar radiation field (ISRF) and the stellar radiation field can contribute to the overall inverse Compton losses experienced by CR electrons. The expression for inverse Compton energy losses from the ISRF follows similarly to \Cref{eq:synch_losses}, 
\begin{equation}
    b_\mathrm{IC, ISRF}(p, \vec{x}) = \frac{4}{3} \sigma_T c \left(\frac{p_e}{m_e c}\right)^2 \epsilon_\mathrm{ph},
\end{equation} 
where energy density contribution from the ISRF can be approximately described by setting $\epsilon_\mathrm{ph} \approx \SI{0.5}{\electronvolt\per\centi\meter\cubed}$ \citep{Mathis1983}. \par

For the inverse Compton losses from the stellar radiation field, we follow the prescription following Eq. (36) in \citet{Khangulyan2014}, where we assume isotropic scattering of target photons. The energy loss depends primarily on the star temperature and radius, as well as the cell position relative to the origin. Here, we assume a star temperature of $\SI{1e4}{\kelvin}$ (Following stellar tracks from \citet{Ekstroem2012} for stars with mass $\SI{20}{\solarmass}$) and a stellar radius of $15 R_\odot$, following typical values from OB stars \citep{Martins2005}.\par

\begin{table*}[t]
    \caption{Maximum kinetic energy and momentum for CR protons and electrons, determined from the intersection of the acceleration timescales and energy loss timescales at the termination shock, used in this work. }
    \label{tab:max_momenta}
    \centering
    \begin{tabular}{c|cccc}
    \hline \hline

        Simulation & $E_\mathrm{max, p}$ / $\si{\giga\electronvolt}$ & $p_\mathrm{max, p}$ / $m_p c$ & $E_\mathrm{max, e}$ / $\si{\giga\electronvolt}$ & $p_\mathrm{max, e}$ / $m_e c$ \\ \hline
        \texttt{ISM-v40-MDiff} & \num{4.29e4} & \num{4.58e4}& \num{792} & \num{1.55e6}\\
        \texttt{ISM-v40-LDiff} & \num{4.32e6} & \num{4.60e6}& \num{8.94e3}& \num{1.75e7}\\
        \texttt{ISM-v40-HDiff} & \num{4.27e2} & \num{4.56e2}& \num{1.13e2}& \num{2.21e5}\\
        \texttt{ISM-v100-MDiff} & \num{8.94e3} & \num{9.53e3}& \num{2.30e2}& \num{4.50e5}\\

    \end{tabular}
    
\end{table*}

At lower energies, Coulomb losses are also dominant for CR electrons. We follow the prescription from \cite{Gould1972b}:
\begin{align}
    b_\mathrm{Coul, e}(p, \vec{x}) = \dfrac{3 \sigma_T n_e(\vec{x}) m_e c^3}{2 \beta_e} &\Bigg[\ln\left(\frac{m_e c^2 \beta_e \sqrt{\gamma_e - 1}}{\hbar \omega_\mathrm{pl}} \right) - \\
    &\log(2)\left(\frac{\beta_e^2}{2} + \frac{1}{\gamma_e}\right) + \frac{1}{2} + \left(\frac{\gamma_e - 1}{4\gamma_e}\right)^2 \Bigg], 
    \label{eq:coul_loss_electrons}
\end{align}
where $\beta_e, \gamma_e$ now represent the Lorentz beta and Lorentz factor for electrons, respectively.

\subsection{Adiabatic Losses for Cosmic Ray Protons and Electrons} \label{subsec:advective_losses}

Both cosmic ray protons and electrons experience adiabatic losses from the advection of the fluid. The advection timescale is derived by taking the ratio of the characteristic length scale to the gas velocity in each cell. We choose the stand-off radius, $R_0$, as determined in \Cref{eq:standoff_radius}, as the characteristic length scale, so that $t_\mathrm{adv} = R_0 / | \vec{v}|$. The adiabatic loss (in momentum space) is then determined by taking the product of the adiabatic loss coefficient with the normalised momentum, i.e. $b_\mathrm{adiab}(p, \vec{x}) = (1 / t_\mathrm{adv}(\vec{x})) \, p$. This term is applied to the calculation of both the cosmic ray proton and electron steady-state spectrum. 

\subsection{Maximal Energy for Exponential Cutoff} \label{subsec:maximal_energy}

The maximum momenta for the exponential cutoff are determined when the acceleration timescale, $t_\mathrm{accel}$, is equal to either the diffusion timescale, $t_\mathrm{CR, diff} = R_0^2 / \kappa_{\parallel, \mathrm{10}}^2$, or the cooling timescale, considering all energy loss processes as described in \Cref{subsec:energy_loss_prot} and \Cref{subsec:energy_loss_ele} for CR protons and electrons, respectively. \par 

The acceleration timescale can be estimated by the rate at which CRs are accelerated with DSA, $t_\mathrm{accel} \sim 20 \kappa(p) / (c v_\mathrm{shock})$ \citep{Drury1983}, where $\kappa(p) \propto \kappa_{\parallel, \mathrm{10}} p^{-0.5}$ is the momentum-dependence of CR diffusion \citep{Berezinskii1990}. The maximal energy is determined at the intersection between the acceleration timescale and the loss timescale, for each model. \par 

In our current simulation setup, we do not communicate the spectral information between each timestep, and as such do not have information regarding their injection site. Considering that the termination shock is a constant and efficient particle accelerator (c.f. \Cref{fig:crshockparams_term}), we use the average value of the shock parameters at the termination shock for all models for all positions and times. The maximal normalised momenta is then evaluated by taking $p_\mathrm{max, s} = p(E_\mathrm{max, s}) / (m_s c)$. \par 

\Cref{tab:max_momenta} shows the maximum kinetic energy and momentum determined from this method for each simulation, for CR protons and electrons, respectively. We note that the choice to use the maximum momenta at the termination shock yields in $\sim 10\%$ of the total energy budget to be attributed towards high-energy cosmic ray protons ($> \SI{1e4}{\giga\electronvolt}$).

    \end{appendix}

\end{document}